\documentclass[acmtog,screen,nonacm]{acmart}
\usepackage{subcaption}
\usepackage{hyperref}
\usepackage{enumitem}
\usepackage[ruled]{algorithm2e} 
\usepackage{pdfpages} 

\SetAlCapSkip{0.5em}
\SetAlCapHSkip{0pt}
 \usepackage{multirow}

\newcommand{\imgref}[1]{\autoref{#1}}

\newcommand{\TT}[1]{#1} 

\newif\ifshowdiff
\showdifffalse  

\ifshowdiff
  \usepackage{soul}
  \usepackage{xcolor}
  \newcommand{\del}[1]{{\color{red}\st{#1}}}

\else
  \newcommand{\del}[1]{}

\fi

\copyrightyear{2026}
\acmYear{2026}
\setcopyright{acmlicensed}
\acmDOI{XXXXXXX.XXXXXXX}
\acmSubmissionID{806}

\begin{document}
\title[MAS-PNCG]{An Efficient Multilevel Preconditioned Nonlinear Conjugate Gradient Method for Incremental Potential Contact}

\author{Yu Zhang}
\authornotemark[1]
\affiliation{%
  \institution{S-Lab, Nanyang Technological University}
  \country{Singapore}
}
\email{yucrazing@gmail.com}

\author{Xing Shen}
\authornote{Equal contributions.}
\affiliation{%
  \institution{Shanghai AI Laboratory}
  \country{China}
}
\email{shenxing@pjlab.org.cn}

\author{Kemeng Huang}
\authornote{Corresponding author.} 
\affiliation{%
  \institution{University of Hong Kong}
  \country{China}
}
\email{kmhuang@connect.hku.hk}

\author{Wei Chen}
\affiliation{%
  \institution{Zhejiang University}
  \country{China}
}
\email{weichen12@zju.edu.cn}

\author{Yin Yang}
\affiliation{%
  \institution{University of Utah}
  \country{United States of America}
}
\email{yangzzzy@gmail.com}

\author{Taku Komura}
\affiliation{%
  \institution{University of Hong Kong}
  \country{China}
}
\email{taku@cs.hku.hk}

\author{Tiantian Liu}
\affiliation{%
  \institution{Independent Researcher}
  \country{China}
}
\email{ltt1598@gmail.com}

\author{Xingang Pan}
\affiliation{%
  \institution{S-Lab, Nanyang Technological University}
  \country{Singapore}
}
\email{xingang.pan@ntu.edu.sg}

\begin{abstract} 
Incremental Potential Contact (IPC) guarantees intersection-free simulation but suffers from high computational costs due to the expensive Hessian assembly and linear solves required by Newton’s method. While Preconditioned Nonlinear Conjugate Gradient (PNCG) avoids Hessian assembly, it has historically struggled with poor convergence in stiff, contact-rich scenarios due to the lack of effective preconditioners; simple Jacobi preconditioners fail to capture the global coupling, while advanced hierarchy-based preconditioners like Multilevel Additive Schwarz (MAS) are computationally prohibitive to rebuild at every nonlinear iteration.
We present MAS-PNCG, a method that unlocks the power of hierarchical preconditioning for nonlinear optimization. Our key technical innovation is a Sparse-Input Woodbury update algorithm that incrementally adapts the fine-level MAS components to rapidly evolving contact sets. This bypasses the need for full preconditioner rebuilds, reducing maintenance cost to near-zero while capturing the complex spectral properties of the contact system. Furthermore, we replace heuristic PNCG search directions with a Hessian-aware 2D subspace minimization that optimally combines the preconditioned gradient and previous direction. We also apply a fast per-subdomain conservative CCD method that ensures penetration-free trajectories while avoiding overly restrictive global step sizes. Experiments demonstrate that our MAS-PNCG outperforms state-of-the-art Newton-PCG solvers, GIPC and StiffGIPC, both preconditioned with MAS up to 5.66$\times$ and 2.07$\times$ respectively.

\end{abstract}


\begin{teaserfigure}
\centering
\includegraphics[width=0.9\textwidth]{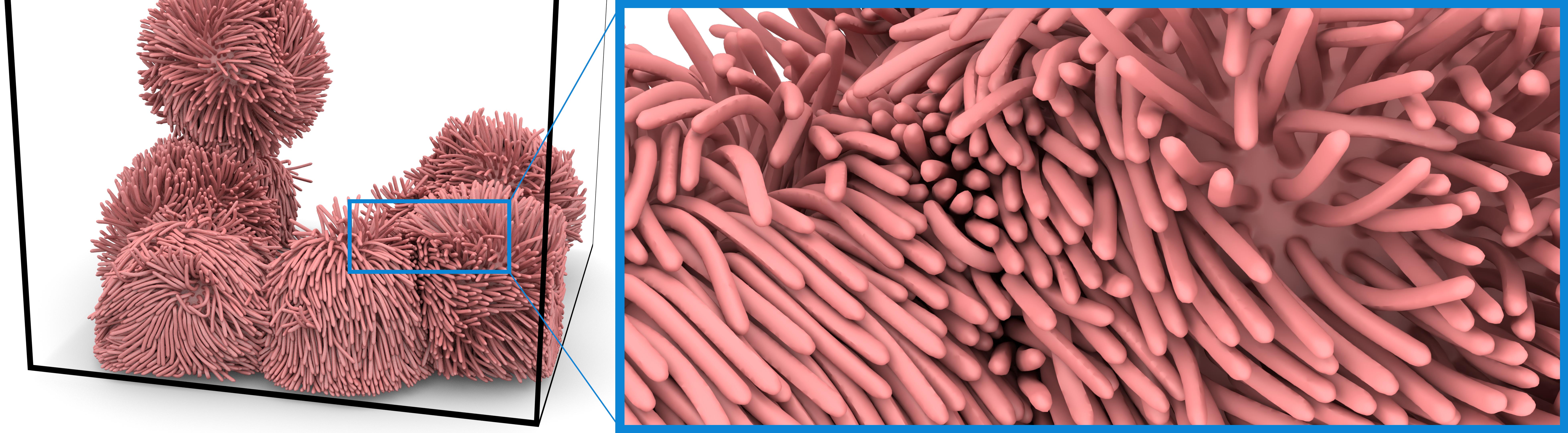}
\captionof{figure}{
Eight puffer balls are dropped from mid-air and collide repeatedly.  
The scene contains more than $1.14\,$M vertices, $2.72\,$M tetrahedra, and over $1\,$M peak active contacts. The high degree of freedom and rapidly changing contact sets challenge classical Jacobi-PNCG, leading to both prohibitive runtimes and penetration issues.  
In contrast, our solver advances efficiently ($99$ seconds / frame), and attains convergence quality comparable to a full Newton method. The zoom-in view shows that our method is able to maintain accurate penetration-free result even in highly complex large-scale collision scenarios.
}
\label{pics:teaser}
\end{teaserfigure}


\maketitle

\section{INTRODUCTION}

Contact handling lies at the heart of physically-based simulation. The faithful resolution of collision determines whether a simulation looks realistic or suffers from visible artifacts like interpenetration and jittering. The Incremental Potential Contact (IPC) method~\cite{li2020incremental,harmon2009asynchronous} marked a significant advance by reformulating contact dynamics as barrier-based unconstrained optimization, guaranteeing intersection-free results while maintaining high accuracy. This approach has since been adopted across diverse applications including multi-material simulations~\cite{li2024dynamic,xie2023contact}, interlocked rigid components~\cite{tang2023beyond}, high-order meshes~\cite{ferguson2023high}, sticky interactions~\cite{fang2023augmented}, and differentiable simulation~\cite{huang2024differentiable}.

However, IPC's reliance on Newton-type solvers~\cite{hecht2012updated,gast2015optimization} creates a computational bottleneck. At each iteration, Newton methods must assemble the full Hessian matrix
and then solve a linear system involving it. For large-scale scenarios with thousands of contact pairs, this becomes prohibitively expensive. Existing acceleration strategies face a fundamental tension: approximation-based methods~\cite{lan2021medial,lan2022penetration,lan2023second,li2023subspace} sacrifice IPC's accuracy for speed, while exact GPU-accelerated Newton solvers like GIPC~\cite{huang2024gipc} and StiffGIPC~\cite{huang2025stiffgipc} preserve accuracy but remain bound to the cost of Hessian assembly and linear solves at each Newton iteration.

Preconditioned Nonlinear Conjugate Gradient (PNCG)~\cite{shen2024preconditioned} offers an alternative that sidesteps Hessian assembly entirely, using only gradient evaluations to iteratively descend toward the energy minimum. This yields dramatically lower per-iteration costs and naturally handles degenerate configurations. However, existing PNCG implementations for IPC rely on simple preconditioners like the Jacobi matrix, which fail to capture the complex coupling between vertices induced by stiff materials and dense contacts. The result is slow convergence that negates the per-iteration savings.

A natural idea is to employ a more powerful preconditioner. The Multilevel Additive Schwarz (MAS) preconditioner~\cite{wu2022gpu}, which combines local domain decomposition with hierarchical coarsening, has proven highly effective for accelerating linear solvers in elastodynamics and was recently adopted by StiffGIPC~\cite{huang2025stiffgipc} for IPC's Newton inner loop. However, applying MAS to nonlinear PNCG faces a fundamental obstacle: MAS requires building subdomain factorizations from the system Hessian, yet PNCG deliberately avoids Hessian assembly. In Newton-PCG methods like StiffGIPC, this is not a problem because the Hessian is fixed within each Newton iteration, allowing a single MAS build to serve all inner PCG steps. But in nonlinear PNCG, the effective curvature changes \emph{every iteration} as contacts evolve---new pairs activate when objects approach within the threshold $\hat{d}$, existing contacts stiffen as gaps shrink, and contacts release as objects separate. Rebuilding the full MAS hierarchy at each iteration would be prohibitively expensive, defeating the purpose of avoiding Newton's Hessian assembly.

Our key observation is that while IPC's effective Hessian changes rapidly, \emph{how} it changes has exploitable structure. Each contact contributes a localized stiffness term that depends on just a few vertices (typically 4--8) and is dominated by a rank-1 component along the contact normal direction \cite{huang2024gipc}. Physically, this reflects the fact that contact forces primarily resist motion along the direction of potential penetration. This sparse, low-rank structure suggests a fundamentally different strategy: instead of rebuilding the MAS preconditioner from scratch at each iteration, we can \emph{incrementally update} the fine-level subdomain inverses through efficient rank-1 Woodbury modifications that track contact stiffening and release, while keeping the coarse levels frozen.

Building on this insight, we develop MAS-PNCG. The coarse MAS levels, which capture global elastic coupling, are built once and updated only periodically. The fine-level subdomain inverses, which are most sensitive to local contact changes, are maintained through Sparse-Input Woodbury updates derived directly from barrier Hessian structure. Since these updates provide approximate curvature information, we can move beyond heuristic $\beta$ formulas to optimal 2D subspace minimization---directly computing the best combination of preconditioned gradient and previous direction. Finally, rather than restricting the entire simulation to a single conservative step size dictated by the stiffest contact, we introduce per-subdomain CCD that allows different regions to advance at locally appropriate rates while maintaining penetration-free guarantees.


Our contributions are:
\begin{itemize}
    \item \TT{\textbf{Incremental Spectral Maintenance via Sparse-Input Woodbury}:
    We introduce an algorithm to update fine-level MAS subdomain inverses using rank-1 Woodbury modifications. This allows the preconditioner to evolve with the contact constraints at negligible cost, solving the bottleneck that previously prevented the use of hierarchical preconditioners in nonlinear CG.}
    \item \TT{\textbf{Curvature-Optimal Subspace Minimization}: We replace the heuristic $\beta$ formulas of classical PNCG with a principled minimization strategy. By projecting the local Hessian approximation into the subspace spanned by the gradient and search direction, we analytically compute the optimal descent step, significantly improving convergence rates in stiff scenarios.}
    \item \textbf{Conservative CCD with Tighter Lower Bounds}: We apply a per-subdomain collision detection scheme that maintains penetration-free guarantees while avoiding global step size restrictions, with frame-invariant lower bounds for improved tightness.
\end{itemize}

Experiments demonstrate that MAS-PNCG achieves up to 5.66$\times$ and 2.07$\times$ speedup over state-of-the-art GIPC~\cite{huang2024gipc} and StiffGIPC~\cite{huang2025stiffgipc} respectively, while converging to comparable accuracy in complex, large-scale scenarios. Moreover, since both GIPC and StiffGIPC already employ MAS preconditioning in their inner PCG solvers, our speedups demonstrate that the performance gains stem not from MAS alone, but from the synergy of our Sparse-Input Woodbury updates and 2D subspace minimization strategy---techniques that enable efficient preconditioner maintenance and enhanced per-iteration progress within the nonlinear CG framework.

\section{RELATED WORK}
\label{sec:related_work}


\subsection{IPC Solvers}

The IPC framework~\cite{li2020incremental} reformulates contact as barrier-based optimization, enabling intersection-free simulation across diverse applications~\cite{li2024dynamic,xie2023contact,tang2023beyond,ferguson2023high,fang2023augmented,huang2024differentiable}. However, its expensive solve cost, especially in Hessian assembly and linear system solve per iteration, severely limits its practical applications.

To accelerate IPC, recent approximation-based methods trade accuracy for efficiency, such as medial axis reduction~\cite{lan2021medial}, projective dynamics~\cite{lan2022penetration}, stencil-wise descent~\cite{lan2023second} and subspace methods~\cite{li2023subspace}. In contrast, our method aims to deliver the acceleration while preserving accuracy comparable to original Newton solver based IPC method.

Accuracy-preserving methods such as GIPC~\cite{huang2024gipc} and its enhanced version StiffGIPC~\cite{huang2025stiffgipc} accelerate IPC through GPU parallelization. For these methods, however, Hessian assembly and linear solves remain the primary computational bottleneck per Newton iteration. Instead, inspired by MAS~\cite{wu2022gpu}, we design a convergence-enhanced nonlinear solver with a tailored multilevel preconditioner. Our approach avoids Hessian assembly entirely by using gradient-only iterations with incrementally updated preconditioners.

\subsection{Nonlinear Conjugate Gradient Methods}

Nonlinear CG methods have been applied to graphics problems~\cite{wang2016descent}. Shen et al.~\cite{shen2024preconditioned} applied PNCG to IPC with Jacobi preconditioning, but observed poor convergence in stiff scenarios. We address this through MAS preconditioning with Woodbury updates and curvature-aware subspace minimization.

Quasi-Newton methods like L-BFGS~\cite{nocedal1999numerical} approximate Hessian information through gradient differences~\cite{al2014damped}. In contrast, our Sparse-Input Woodbury approach exploits IPC's structure where contact contributions are sparse and low-rank, deriving updates directly from barrier derivatives.

\subsection{Multilevel Additive Schwarz Preconditioning}

Domain decomposition methods~\cite{toselli2004domain,smith1996domain} solve large sparse linear systems by partitioning domains into overlapping subdomains. The Multilevel Additive Schwarz variant adds coarse-level corrections for faster convergence.

Wu et al.~\cite{wu2022gpu} introduced GPU-parallelized MAS for elastodynamics within Newton solvers. StiffGIPC~\cite{huang2025stiffgipc} adapted this for IPC, applying MAS to the linear PCG inner loop.

A crucial distinction separates these prior applications from ours. In Newton methods, MAS preconditions a \emph{fixed} linear system $\mathbf{H}\mathbf{d} = -\mathbf{g}$ within each iteration---the Hessian and MAS hierarchy remain constant throughout the inner solve. In contrast, we apply MAS to nonlinear PNCG where effective curvature changes every iteration as contacts evolve. The challenge is maintaining preconditioner quality across rapidly changing landscapes without explicit Hessian matrices. We address this through Sparse-Input Woodbury updates exploiting the sparse, low-rank structure of contact barrier Hessians.

\subsection{Continuous Collision Detection}

CCD ensures intersection-free trajectories by computing time-of-impact before penetration. Recent robust methods include conservative advancement~\cite{li2021codimensional}, inclusion-based root-finding~\cite{wang2021large}, exact root parity~\cite{wang2022fast}, and GPU-scalable algorithms~\cite{belgrod2021time}.

Standard CCD computes a global step size, causing one stiff contact to restrict motion everywhere. Our per-subdomain approach computes localized step sizes, allowing different regions to advance appropriately while maintaining penetration-free guarantees.

\section{BACKGROUND}
\label{sec:background}


\subsection{Incremental Potential Contact}

IPC formulates contact simulation as minimizing:
\begin{equation}
    E(\mathbf{x}) = \underbrace{\frac{1}{2}(\mathbf{x}-\tilde{\mathbf{x}})^{\top} \mathbf{M}(\mathbf{x}-\tilde{\mathbf{x}})}_{\text{inertia}} + \underbrace{h^2 \Psi(\mathbf{x})}_{\text{elasticity}} + \underbrace{B(\mathbf{x})}_{\text{contact barrier}} + \underbrace{D(\mathbf{x})}_{\text{friction}},
    \label{eq:ipc_energy}
\end{equation}
where $\tilde{\mathbf{x}} = \mathbf{x}^{t} + h\mathbf{v}^{t} + h^2\mathbf{M}^{-1}\mathbf{f}_{\text{ext}}$ incorporates previous state and external forces. The contact barrier $B(\mathbf{x}) = \kappa \sum_{c \in \mathcal{C}} b(d_c(\mathbf{x}))$ activates when distance $d < \hat{d}$, with $b(d)$ a logarithmic function ensuring non-penetration.

\subsection{Preconditioned Nonlinear Conjugate Gradient}

PNCG generates iterates $\mathbf{x}_{k+1} = \mathbf{x}_k + \alpha_k \mathbf{p}_{k+1}$, where the search direction
\begin{equation}
    \mathbf{p}_{k+1} = -\mathbf{P}_{k+1} \mathbf{g}_{k+1} + \beta_k \mathbf{p}_{k}
    \label{eq:pncg_direction}
\end{equation}
combines the preconditioned gradient $\mathbf{P}_{k+1} \mathbf{g}_{k+1}$ with the previous direction $\mathbf{p}_k$ via conjugate parameter $\beta_k$.

\subsection{Multilevel Additive Schwarz Preconditioner}

MAS combines domain decomposition with hierarchical coarsening. The domain is partitioned into $D$ subdomains with selection matrices $\mathbf{S}_d$. The level-0 preconditioner aggregates local inverses:
\begin{equation}
    \mathbf{M}_{(0)}^{-1} = \sum_{d=1}^D \mathbf{S}_d^T (\mathbf{S}_d \mathbf{A} \mathbf{S}_d^T)^{-1} \mathbf{S}_d.
\end{equation}
The full MAS preconditioner extends this hierarchically:
\begin{equation}
    \mathbf{P} = \mathbf{M}_{(0)}^{-1} + \sum_{l=1}^L \mathbf{C}_{(l)}^T \mathbf{M}_{(l)}^{-1} \mathbf{C}_{(l)},
\end{equation}
where $\mathbf{C}_{(l)}$ are coarsening matrices.

\section{METHOD}
\subsection{Localized Woodbury Level-0 Update}
The MAS preconditioner significantly accelerates NCG convergence~\cite{wu2022gpu}, yet frequent full recomputations remain computationally prohibitive. 
\TT{Our method inherits the setup of MAS in StiffGIPC~\cite{huang2025stiffgipc}, the details can be found in the supplemental document.}
Our update strategy leverages the spectral properties of the system: coarse-level components primarily capture low-frequency error modes, which evolve relatively slowly. Consequently, these components can be temporarily frozen without significantly degrading convergence.

Furthermore, Hessian contributions from elastic potential energy tend to remain stable or change smoothly. In contrast, severe ill-conditioning arises primarily from contact barrier terms, which activate and deactivate abruptly. Fortunately, these interactions are spatially sparse, affecting only a small subset of the domain. Therefore, we focus our update strategy exclusively on the Level-0 (fine-level) subdomains to capture these local changes via efficient low-rank updates, while keeping the rest of the preconditioner fixed.

{
To ensure the preconditioner remains Symmetric Positive Definite (SPD), we note that the contact barrier Hessian has positive eigenvalues only in the normal direction, tangential eigenvalues are negative and thus eliminated by SPD projection.} With scalar stiffness $k = b''(d_c)$ and contact normal $\mathbf{n} = \nabla d_c$, the projected Hessian reduces to:
\begin{equation}
    \nabla^2 b(d_c(\mathbf{x})) \approx k \mathbf{n} \mathbf{n}^T.
\end{equation}
This simplification allows us to model the Hessian change as a sum of rank-1 updates. 
We define the base state using the frozen Hessian snapshots $\mathbf{H}_{\text{base}}$ 
and inverse of each subdomain of level 0 $\mathbf{B}_d = (\mathbf{M}_{0}^d)^{-1}$ from the last global restart. The current Hessian is then approximated as:
\begin{equation}\label{eq: Hessian}
    \tilde{\mathbf{H}} = \mathbf{H}_{\text{base}} + \mathbf{U} \mathbf{U}^T,
\end{equation}
where $\mathbf{U} = [\mathbf{u}_1, \mathbf{u}_2, \dots, \mathbf{u}_m]$ is the update matrix collecting the rank-1 contributions $\mathbf{u}_i = \sqrt{k_i} \mathbf{n}_i$ from $m$ active contacts. Since the update takes the form $\mathbf{U}\mathbf{U}^T$, the Woodbury formula simplifies with an identity coupling matrix. This formulation enables efficient low-rank updates to the Level-0 preconditioner for capturing rapidly changing contact states, as detailed below.

\subsubsection{Construction of Local Update Matrices}
We maintain a global set of active contacts $\mathcal{C}_{\text{curr}}$ and compare each contact $c \in \mathcal{C}_{\text{curr}}$ against its state in the global base set $\mathcal{C}_{\text{base}}$ from the last restart. Using a rotation threshold $\epsilon_{\text{rot}}$ (e.g., $25^\circ$), we determine the potential rank-1 update vectors $\mathbf{u}$ and their significance metrics $\Delta S$ by classifying each contact into two categories:

\begin{enumerate}[leftmargin=*]
    \item \textbf{Case 1: New Contact} ($c \notin \mathcal{C}_{\text{base}}$).
    For topologically new contacts, we inject the full stiffness as a pure increment:
    \begin{equation}
        \mathbf{u} = \sqrt{k_{\text{curr}}} \cdot \mathbf{n}_{\text{curr}}, \quad \Delta S = k_{\text{curr}}.
    \end{equation}

    \item \textbf{Case 2: Existing Contact} ($c \in \mathcal{C}_{\text{base}}$).
    For contacts persisting from the base state, we categorize them based on the stability of their normal directions.
    \begin{itemize}[leftmargin=*, label={$\bullet$}]
        \item \textbf{Rotated Normal ($\mathbf{n}_{\text{curr}} \cdot \mathbf{n}_{\text{base}} < \epsilon_{\text{rot}}$):}
        If the geometric constraint has rotated significantly, updating the direction is numerically unstable. We treat this as a \textbf{New Contact}, effectively retaining the old stiffness as a "ghost wall" while adding the new constraint to ensure robustness:
        \begin{equation}
            \mathbf{u} = \sqrt{k_{\text{curr}}} \cdot \mathbf{n}_{\text{curr}}, \quad \Delta S = k_{\text{curr}}.
        \end{equation}

        \item \textbf{Stable Normal ($\mathbf{n}_{\text{curr}} \cdot \mathbf{n}_{\text{base}} \ge \epsilon_{\text{rot}}$):}
        For directionally stable contacts, we further compare the stiffness:
        \begin{itemize}[leftmargin=*, label={$-$}]
            \item \textit{Stiffening ($k_{\text{curr}} > k_{\text{base}}$):} We add the stiffness differential:
            \begin{equation}
                \mathbf{u} = \sqrt{k_{\text{curr}} - k_{\text{base}}} \cdot \mathbf{n}_{\text{curr}}, \quad \Delta S = k_{\text{curr}} - k_{\text{base}}.
            \end{equation}
            \item \textit{Softening ($k_{\text{curr}} \le k_{\text{base}}$):} We ignore stiffness decreases. Retaining the stiffer base approximation yields a safer step without risking indefiniteness. We set $\Delta S = 0$.
        \end{itemize}
    \end{itemize}
\end{enumerate}

To balance approximation accuracy and computational overhead, we employ a \textit{localized Top-K truncation} strategy. For each subdomain $d$, we rank all potential local update vectors by their significance metric $\Delta S$ and select only the top $K$ vectors (e.g., $K=8$) to form the subdomain update matrix $\mathbf{U}_d \in \mathbb{R}^{N_d \times K}$. This ensures the rank of the update remains low within each subdomain, keeping the subsequent per-subdomain capacitance solve efficient.

\subsubsection{Localized Woodbury Solver}
To maintain the efficiency of the domain-decomposed preconditioner without full rebuilds, we apply the Woodbury update locally within each subdomain $d$. Let $\mathbf{M}_{\text{base}}^d$ be the Level-0 subdomain Hessian frozen at the last global restart. We factorize $\mathbf{M}_{\text{base}}^d$ and cache its inverse $\mathbf{B}_d = (\mathbf{M}_{\text{base}}^d)^{-1}$ as the \textit{base state}. The updated inverse $\tilde{\mathbf{B}}_d$ is then efficiently computed via the Woodbury formula:
\begin{equation}
    \tilde{\mathbf{B}}_d = \mathbf{B}_d - \mathbf{B}_d \mathbf{U}_d \left( \mathbf{I}_K + \mathbf{U}_d^T \mathbf{B}_d \mathbf{U}_d \right)^{-1} \mathbf{U}_d^T \mathbf{B}_d.
\end{equation}

For a local gradient $\mathbf{g}_d = \mathbf{S}_d \mathbf{g}$, the preconditioned direction $\mathbf{z}_d = \tilde{\mathbf{B}}_d \mathbf{g}_d$ is computed through a streamlined \textit{capacitance system}:
\begin{enumerate}[leftmargin=*]
    \item \textbf{Base Solve}: Compute the cached projection $\mathbf{z}_{\text{base}} = \mathbf{B}_d \mathbf{g}_d$.
    \item \textbf{Update Projection}: Form the auxiliary matrix $\mathbf{W}_d = \mathbf{B}_d \mathbf{U}_d$ and the correction vector $\mathbf{r} = \mathbf{U}_d^T \mathbf{z}_{\text{base}}$.
    \item \textbf{Capacitance Solve}: Solve the $K \times K$ system $(\mathbf{I}_K + \mathbf{U}_d^T \mathbf{W}_d) \boldsymbol{\lambda} = \mathbf{r}$.
    \item \textbf{Correction}: Update the final local direction $\mathbf{z}_d = \mathbf{z}_{\text{base}} - \mathbf{W}_d \boldsymbol{\lambda}$.
\end{enumerate}

Finally, the global preconditioned gradient $\mathbf{z}_{k+1}$ is assembled by merging the updated Level-0 contributions with the frozen coarse-level corrections:
\begin{equation}
    \mathbf{z}_{k+1} = \sum_{d=1}^D \mathbf{S}_d^T \mathbf{z}_d + \sum_{l=1}^L \mathbf{C}_{(l)}^T \mathbf{M}_{(l)}^{-1} \mathbf{C}_{(l)} \mathbf{g}_{k+1},
\end{equation}
where $\mathbf{M}_{(l)}^{-1}$ are cached from the last restart. 
\paragraph{Efficiency Optimization}
The localized Woodbury solver provides a critical advantage: it completely bypasses the need for global Hessian assembly and expensive matrix factorization in each iteration. By reusing the frozen base inverse $\mathbf{B}_d$ and only updating it with a low-rank matrix $\mathbf{U}_d$, we maintain an up-to-date preconditioner at minimal cost. This effectively enables the preconditioner to "evolve" with the contact state without a full rebuild, ensuring both numerical stability and high performance.

\begin{algorithm}[htbp!]
\small
\caption{MAS-PNCG Pipeline}\label{alg:MAS}
\KwIn{Position $\mathbf{x}^t$, velocity $\mathbf{v}^t$, mass $\mathbf{M}$, barrier $\hat{d}$, tol $\epsilon$, restart $\delta$}
\KwOut{Updated $\mathbf{x}^{t+1}$}
$\mathbf{x}_0 \gets \mathbf{x}^{t}$;\quad $\tilde{\mathbf{x}} \gets \mathbf{x}^{t} + h\mathbf{v}^{t} + h^2\mathbf{M}^{-1}\mathbf{f}_{\text{ext}}$;\quad $\text{Restart} \gets \texttt{True}$\;
\For{$k = 0$ \KwTo $\text{IterMax}$}{
    $\mathcal{C} \gets \textsc{ComputeConstraintSet}(\mathbf{x}_k, \hat{d})$\;
    \uIf{$\text{Restart}$}{
        $(\mathbf{P}_{\text{base}}, \tilde{\mathbf{H}}) \gets \textsc{RebuildMAS}(\mathbf{x}_{k}, \mathcal{C})$;\quad $\mathbf{P}_{k+1} \gets \mathbf{P}_{\text{base}}$\;
    }\Else{
        $\mathbf{P}_{k+1} \gets \textsc{WoodburyUpdate}(\mathbf{P}_{\text{base}}, \mathcal{C})$\;
    }
    $\mathbf{g}_{k+1} \gets \nabla E(\mathbf{x}_k, \tilde{\mathbf{x}}, \mathcal{C})$;\quad $\mathbf{z}_{k+1} \gets \mathbf{P}_{k+1} \mathbf{g}_{k+1}$;\quad $\mathbf{v} \gets \tilde{\mathbf{H}} \mathbf{z}_{k+1}$\;
    \uIf{$\text{Restart}$}{
        $\mu \gets (\mathbf{z}_{k+1}^\top \mathbf{g}_{k+1}) / (\mathbf{z}_{k+1}^\top \mathbf{v})$;\quad $\nu \gets 0$\;
    }\Else{
        $(\mu, \nu) \gets \textsc{Solve2DSubspace}(\mathbf{z}_{k+1}, \mathbf{p}_k, \mathbf{g}_{k+1}, \tilde{\mathbf{H}})$
    }
    $\mathbf{p}_{k+1} \gets -\mu \mathbf{z}_{k+1} + \nu \mathbf{p}_k$\;
    $\{\alpha_d\} \gets \textsc{ConservativeCCD}(\mathbf{x}_k, \mathbf{p}_{k+1})$\;
    $\mathbf{x}_{k+1} \gets \mathbf{x}_{k} + \sum_d \mathbf{S}_d^T \alpha_d \mathbf{S}_d \mathbf{p}_{k+1}$\;
    \If{$\|\mathbf{z}_{k+1}\| \leq \epsilon$ \textbf{and}  $\text{Restart }$}{
        \textbf{break}\;
    }
    $r_k \gets |\mathbf{g}_{k+1}^\top \mathbf{z}_k| / (\mathbf{g}_{k+1}^\top \mathbf{z}_{k+1})$\;
    $\text{Restart} \gets (r_k > \delta)$\;
    $\mathbf{z}_k \gets \mathbf{z}_{k+1}$\;
}
\Return{$\mathbf{x}_{k+1}$}\;
\end{algorithm}

\subsection{Optimal 2D Subspace Minimization}
\label{sec:subspace_minimization}

Traditional PNCG methods typically lack direct access to Hessian information, relying instead on heuristic conjugacy parameters (e.g., Polak-Ribière) and decoupled line searches. Leveraging our approximated Hessian (Eq.~\ref{eq: Hessian}), we can directly determine the optimal search direction $\mathbf{p}_{k+1}$ within the 2D subspace spanned by the preconditioned gradient $\mathbf{z}_{k+1} = \mathbf{P}_{k+1} \mathbf{g}_{k+1}$ and the previous direction $\mathbf{p}_k$:
\begin{equation}
    \mathbf{p}_{k+1}(\mu, \nu) = -\mu \mathbf{z}_{k+1} + \nu \mathbf{p}_k.
    \label{eq:subspace_dir}
\end{equation}
The optimal coefficients $(\mu^*, \nu^*)$ are obtained by minimizing the local quadratic model $Q(\mathbf{p}_{k+1}) = \mathbf{g}_{k+1}^\top \mathbf{p}_{k+1} + \frac{1}{2} \mathbf{p}_{k+1}^\top \tilde{\mathbf{H}} \mathbf{p}_{k+1}$, which yields the following $2 \times 2$ system:
\begin{equation}
    \begin{bmatrix} 
    \mathbf{z}_{k+1}^\top \tilde{\mathbf{H}} \mathbf{z}_{k+1} & -\mathbf{z}_{k+1}^\top \tilde{\mathbf{H}} \mathbf{p}_k \\ 
    -\mathbf{p}_k^\top \tilde{\mathbf{H}} \mathbf{z}_{k+1} & \mathbf{p}_k^\top \tilde{\mathbf{H}} \mathbf{p}_k 
    \end{bmatrix} 
    \begin{bmatrix} \mu \\ \nu \end{bmatrix} 
    = 
    \begin{bmatrix} \mathbf{z}_{k+1}^\top \mathbf{g}_{k+1} \\ -\mathbf{p}_k^\top \mathbf{g}_{k+1} \end{bmatrix}.
    \label{eq:2x2_system}
\end{equation}
This formulation eliminates the need for heuristic $\beta$ selection and expensive line searches. Notably, the coefficient $\mu$ acts as a "natural step size" that accounts for second-order curvature, allowing us to adopt a unit initial trial step ($\alpha = 1.0$) which is only clamped to enforce non-penetration via our Conservative CCD (Section~\ref{sec:conservative_ccd}). 
Detailed derivations are provided in the Supplementary Document.

\subsection{Global Restart Criterion}
\label{sec:restart}
Although Sparse-Input Woodbury updates enhance the efficiency of managing level $0$ MAS components, approximation errors can accumulate across iterations, leading to a decline in preconditioner quality and deterioration of the conjugate property. To address this, we employ a global restart strategy that monitors the orthogonality loss between the current gradient $\mathbf{g}_{k+1}$ and the previous preconditioned gradient $\mathbf{z}_k$:
\begin{equation}
    r_k = \frac{|\mathbf{g}_{k+1}^\top \mathbf{z}_k|}{\mathbf{g}_{k+1}^\top \mathbf{z}_{k+1}}.
\end{equation}
When $r_k$ exceeds a threshold (e.g., $\delta = 0.3$), indicating significant loss of orthogonality in the preconditioned space, we trigger a full MAS rebuild. Evaluating in the preconditioned space better reflects the true loss of conjugacy, as it filters out the ill-conditioning of the original system.
\vspace{-5pt} 
\subsection{Localized Conservative CCD}
\label{sec:conservative_ccd}

While our subspace minimization strategy provides an optimal trial step, it cannot guarantee collision-free motion in non-quadratic, contact-rich landscapes. To address the "numerical locking" inherent in global CCD, where a single difficult contact restricts the entire system, we implement Conservative CCD (CCCD) by computing safe steps $\{\alpha_d\}$ independently for each subdomain, unconstrained regions maintain optimal descent speeds while only critical subdomains are damped, ensuring that local restrictions do not compromise global convergence.

Our CCCD strategy prioritizes rapid, safe clamping over exact Time-of-Impact (TOI) calculation. The key insight is that penetration detection often signals a search direction contaminated by missing collision constraints. Rather than pursuing expensive exact root-finding to ``touch'' the obstacle, we delegate accuracy to the subsequent iteration—injecting the newly detected contact into the Woodbury update naturally steers the search direction away from the collision.

The CCCD algorithm examines the cubic distance function $f_{c}(\alpha)$ over the interval $[0, 1.0]$. Following~\cite{cypo}, we first identify the monotonic region by computing $\alpha_{\text{mon}}$, the first critical point of $f'_{c}(\alpha)$. Within the restricted interval $[0, \min(1.0, \alpha_{\text{mon}})]$, we employ a sign-based root detection strategy: if $\text{sign}(f_{c}(0)) \neq \text{sign}(f_{c}(\alpha))$, a collision exists within the interval. Rather than precisely locating the collision point, we use a bisection strategy that iteratively halves $\alpha$ until either the signs match (indicating no collision) or we reach the minimal threshold $\alpha_l$. This approach avoids the numerical instabilities of polynomial root-finding, requires only function evaluations, and achieves rapid convergence to safe step sizes while avoiding ACCD’s potentially excessive iterations. 
Furthermore, We introduce an improved lower bound for CCD step sizes by directly bounding relative displacements between primitives, providing a tighter estimate compared to standard
ACCD~\cite{li2021codimensional}. Complete algorithmic details and pseudocode are provided in the Supplementary Document.

\section{EXPERIMENTS AND ANALYSIS}
We conducted all experiments on a desktop workstation with an \texttt{Intel\,Core~i9-13900X} CPU and an \texttt{NVIDIA\,RTX\,4090} GPU.  
Our method is built on top of the publicly available StiffGIPC code base \cite{huang2025stiffgipc}, and all simulations performed using double-precision floating-point arithmetic. 

Table~\ref{tab:performance_summary_combined_simple} presents the experimental settings and overall performance results of MAS-PNCG across diverse simulation scenarios, ranging from the 22K-tetrahedra "Armadillo" (\imgref{pics:armadillo}) to the highly demanding "Puffer Balls" simulation (\imgref{pics:puffer_balls}) with 2.72M tetrahedral elements.  Our method demonstrates consistent performance across various energy models, including As-Rigid-As-Possible (ARAP), Stable Neo-Hookean (SNH) \cite{smith2018stable}, and Baraff-Witkin cloth~\cite{kim2020finite}. 
The average iteration counts scale reasonably with problem complexity, demonstrating our method's versatility while maintaining robust intersection-free guarantees.

\begin{table}[t]
  \centering
  \small
  \caption{Experimental Settings and Overall Performance Results of MAS-PNCG across Various Benchmarks. The "Mesh" column shows vertex/face/tetrahedron counts (faces shown only for cloth). Time step $\mathbf{h}$ is 0.01s for Armadillo (\imgref{pics:armadillo}) through Cloth Twisting (\imgref{pics:cloth_twisting}), and 0.005s for remaining scenes. Gap distance $\mathbf{\hat{d}}=10^{-3}$ except Teapots (\imgref{pics:teapots}) and Dropping Letters (\imgref{pics:sig_asia}) ($5\times10^{-4}$) and Puffer Balls (\imgref{pics:puffer_balls}) ($10^{-4}$). "Avg. Iters" denotes average NCG iterations per frame. "Time" shows average wall-clock time (\textbf{s}) per frame. The abbreviations ARAP, SNH and BW represent As-Rigid-As-Possible, Stable Neo-Hookean and Baraff-Witkin cloth energy models, respectively.}
\label{tab:performance_summary_combined_simple}
\resizebox{0.45\textwidth}{!}{%
  \begin{tabular}{l|c|r|r|r}
    \hline
    \textbf{Example} & \textbf{Mesh (\#V/\#F/\#T)} & \textbf{Material} & \textbf{Avg. Iters} & \textbf{Time} \\
    \hline
    \imgref{pics:armadillo} & 8K / - / 22K & ARAP & 29 & 0.11 \\
    \imgref{pics:octopus_stack} & 10K / - / 39K & ARAP & 36 & 0.25 \\
    \imgref{pics:single_bunny} & 20K / - / 80K & SNH & 26 & 0.13  \\
    \imgref{pics:two_bunnies} & 40K / - / 160K & SNH & 33 & 0.21  \\
    \imgref{pics:cloth_twisting} & 20K / 40K / - & BW & 62 & 3.29\\
    \imgref{pics:bunny_cloth} & 110K / 179K /80K & SNH+BW & 107 & 7.44  \\
    \imgref{pics:teapots} & 39K / - / 121K & ARAP & 184 & 7.91  \\
    \imgref{pics:dragon} & 78K / - / 274K & SNH & 76 & 3.39  \\
    \imgref{pics:sig_asia} & 105K / - / 298K & SNH & 81 & 4.18  \\
    \imgref{pics:puffer_balls} & 1.14M / - / 2.72M & SNH & 891 & 99.62 \\
    \hline
  \end{tabular}
  }

\end{table}

\begin{table*}[htbp]
  \centering
  \setlength{\abovecaptionskip}{3pt}  
  \setlength{\belowcaptionskip}{3pt}  
  \caption{\textbf{Performance Comparison with Timing Breakdown.} We compare the average time per frame (s) for key solver stages against GIPC \cite{huang2024gipc} and StiffGIPC \cite{huang2025stiffgipc}. Both baselines are preconditioned with MAS in their inner PCG solvers. All timing columns show per-frame time in seconds. The breakdown includes gradient and Hessian computation, search direction computation, CCD, and line search. Speedup is calculated as (Baseline / Ours) and shows: vs GIPC, vs StiffGIPC.}
  \label{tab:performance_comparison_gipc}
  \renewcommand{\arraystretch}{0.9}  
  \small
  \resizebox{0.9\textwidth}{!}{%
  \begin{tabular}{l|c|c|ccccc|c}
    \hline
    Example & Mesh DOFs (v, t) & Method & Grad \& Hess & Moving Dir & CCD & Line Search & Total Time & Speedup \\
    \hline
    \multirow{3}{*}{Octopus Stack} & \multirow{3}{*}{10K, 39K}
    & GIPC & 0.118 & 0.626 & 0.042 & 0.128 & 0.92 & \multirow{3}{*}{\textbf{3.71$\times$, 1.97$\times$}} \\
    & & StiffGIPC & 0.049 & 0.281 & 0.032 & 0.092 & 0.49 & \\
    & & Ours & 0.075 & \textbf{0.023} & 0.066 & 0.079 & \textbf{0.25} & \\
    \hline
    \multirow{3}{*}{Single Bunny} & \multirow{3}{*}{20K, 80K}
    & GIPC & 0.111 & 0.298 & 0.016 & 0.050 & 0.49 & \multirow{3}{*}{\textbf{3.64$\times$, 1.54$\times$}} \\
    & & StiffGIPC & 0.052 & 0.099 & 0.011 & 0.032 & 0.21 & \\
    & & Ours & 0.062 & \textbf{0.019} & 0.027 & 0.023 & \textbf{0.13} & \\
    \hline
    \multirow{3}{*}{Two Bunnies} & \multirow{3}{*}{40K, 160K}
    & GIPC & 0.257 & 0.752 & 0.031 & 0.127 & 1.18 & \multirow{3}{*}{\textbf{5.66$\times$, 2.07$\times$}} \\
    & & StiffGIPC & 0.113 & 0.207 & 0.024 & 0.078 & 0.43 & \\
    & & Ours & 0.090 & \textbf{0.026} & 0.035 & 0.054 & \textbf{0.21} & \\
    \hline
    \multirow{3}{*}{Armadillo} & \multirow{3}{*}{8K, 22K}
    & GIPC & 0.033 & 0.133 & 0.013 & 0.038 & 0.22 & \multirow{3}{*}{\textbf{2.12$\times$, 1.03$\times$}} \\
    & & StiffGIPC & 0.014 & 0.051 & 0.011 & 0.028 & 0.11 & \\
    & & Ours & 0.018 & \textbf{0.011} & 0.023 & 0.052 & \textbf{0.11} & \\
    \hline
    \multirow{3}{*}{Dragon High} & \multirow{3}{*}{78K, 274K}
    & GIPC & 1.565 & 7.895 & 0.412 & 0.981 & 11.05 & \multirow{3}{*}{\textbf{3.25$\times$, 1.16$\times$}} \\
    & & StiffGIPC & 0.845 & 2.281 & 0.227 & 0.465 & 3.95 & \\
    & & Ours & 1.091 & \textbf{0.394} & 0.692 & 1.141 & \textbf{3.39} & \\
    \hline
  \end{tabular}
  }
\end{table*}

\subsection{Ablation and Comparative Studies}



\subsubsection*{Preconditioner Efficacy.} 



\begin{figure}[htbp]
\centering
\begin{subfigure}[t]{0.4\textwidth}
\includegraphics[width=\textwidth]{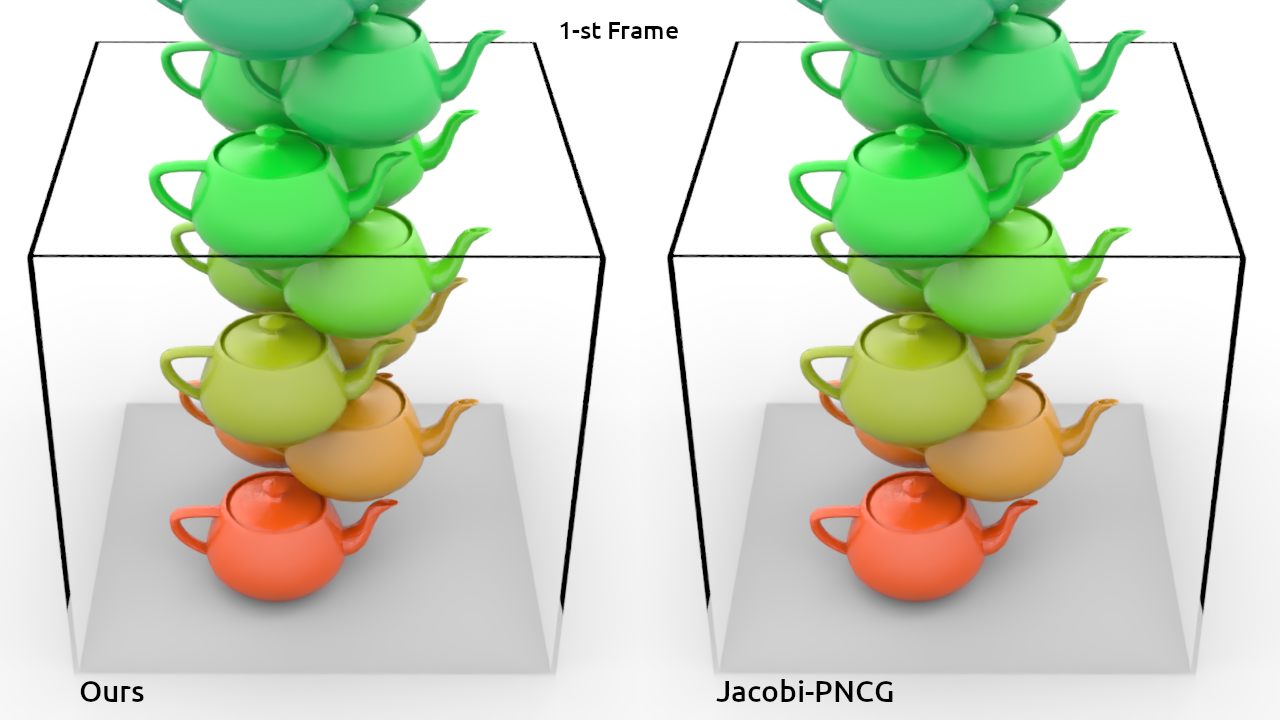}
\end{subfigure}
\begin{subfigure}[t]{0.4\textwidth}
\includegraphics[width=\linewidth]{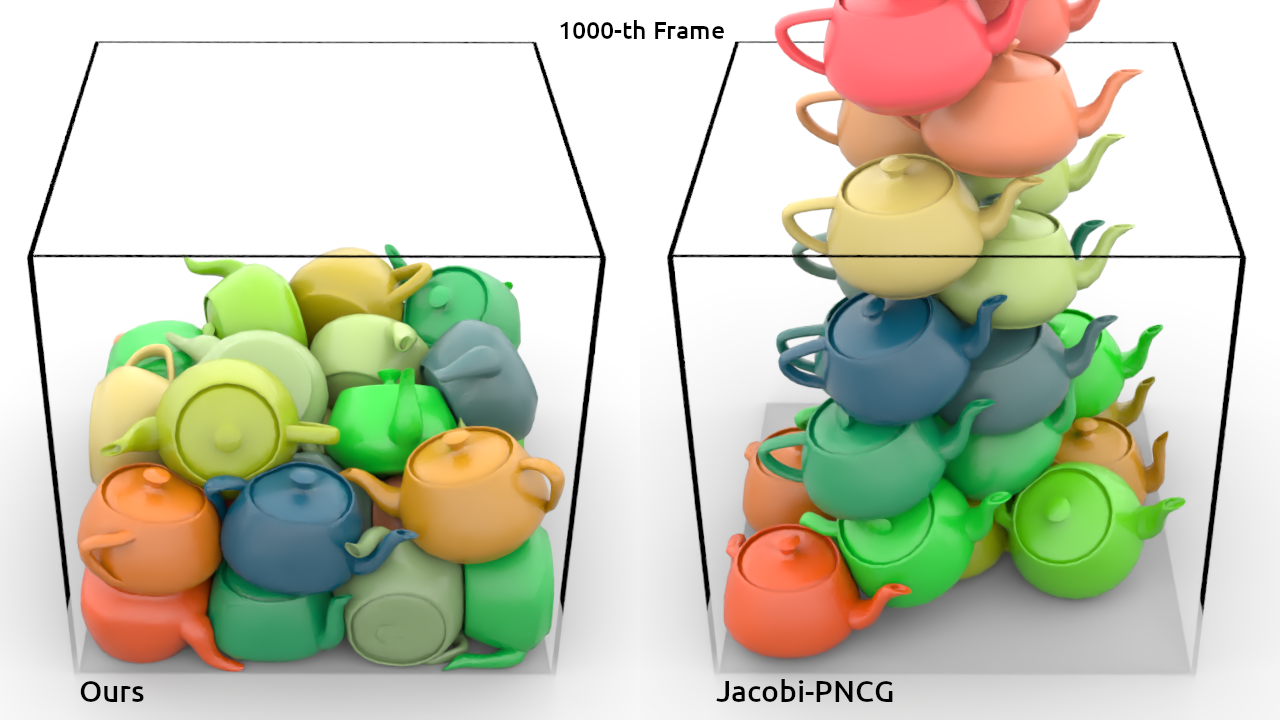}
\end{subfigure}
\vspace{-10pt} 

\caption{\textbf{Teapots.} 32 elastic teapots are dropped into a box. We benchmark the proposed MAS preconditioner against the classical Jacobi preconditioner. For the same wall-clock budget, MAS attains markedly higher accuracy than Jacobi: MAS delivers smooth, physically plausible dynamics, whereas Jacobi exhibits excessive damping and severe distortion—clear symptoms of poor convergence.  In fact, to match the visual fidelity of MAS, Jacobi must run for more than an order of magnitude longer and still occasionally diverges.} 
\vspace{-5pt} 

\label{pics:teapots}

\end{figure}

We benchmark the proposed MAS preconditioner against the classic Jacobi preconditioner in the ``Teapots’’ scene (\imgref{pics:teapots}), where 32 elastic teapots are dropped into a box.  
In this experimenal setting, we use a $3 \times 3$ blocked Jacobi preconditioner. All teapots are simulated with the ARAP energy model and a Young’s modulus of $3\times10^{5}$.  
The convergence tolerance of $\epsilon=1\times10^{-5}$, as looser thresholds frequently lead to noticeable artifacts.

In preliminary experiments, Jacobi-PNCG requires substantially more iterations to reach the same tolerance, leading to prohibitively long runtimes (several minutes per frame) without an iteration cap, while not altering the qualitative conclusions. Therefore, to ensure practical and fair comparisons, we impose a maximum of 10{,}000 PNCG iterations per time step for all solver.

Given the same time budget, MAS achieves significantly higher accuracy than Jacobi. \imgref{pics:teapots} contrasts the two solvers at an intermediate frame: MAS delivers smooth, physically plausible dynamics, whereas Jacobi exhibits excessive damping and severe distortion—clear symptoms of poor convergence. To achieve comparable visual quality, the Jacobi method requires more than ten times longer computation time and may still occasionally fail to converge. 

\vspace{-5pt}
\subsubsection*{Material Stiffness Robustness.}
\begin{figure}[htbp!]
\centering
\begin{subfigure}[b]{0.45\textwidth}
    \centering
    \includegraphics[width=\columnwidth, trim=0 0 0 0, clip]{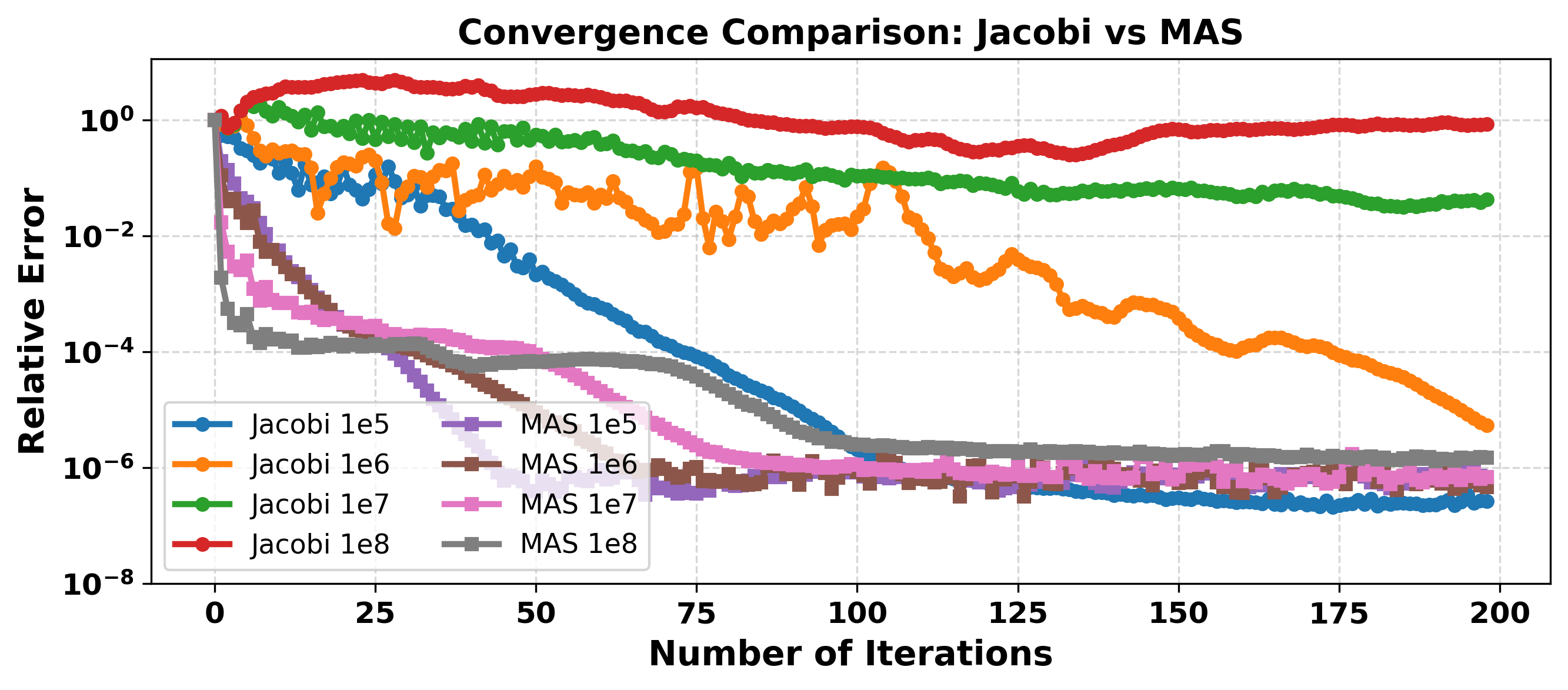}
\end{subfigure}
\vspace{-5pt} 
\caption{Convergence behavior of MAS-PNCG and Jacobi-PNCG under varying material stiffness. MAS-PNCG maintains robust convergence even for highly stiff materials, while Jacobi-PNCG’s convergence degrades severely, failing to reach low error values for high stiffness materials.}
\vspace{-5pt} 
\label{pics:convergence_stiff}
\end{figure}

\begin{figure}[htbp]
\centering
\includegraphics[width=0.465\textwidth, trim=0 40 0 0, clip]{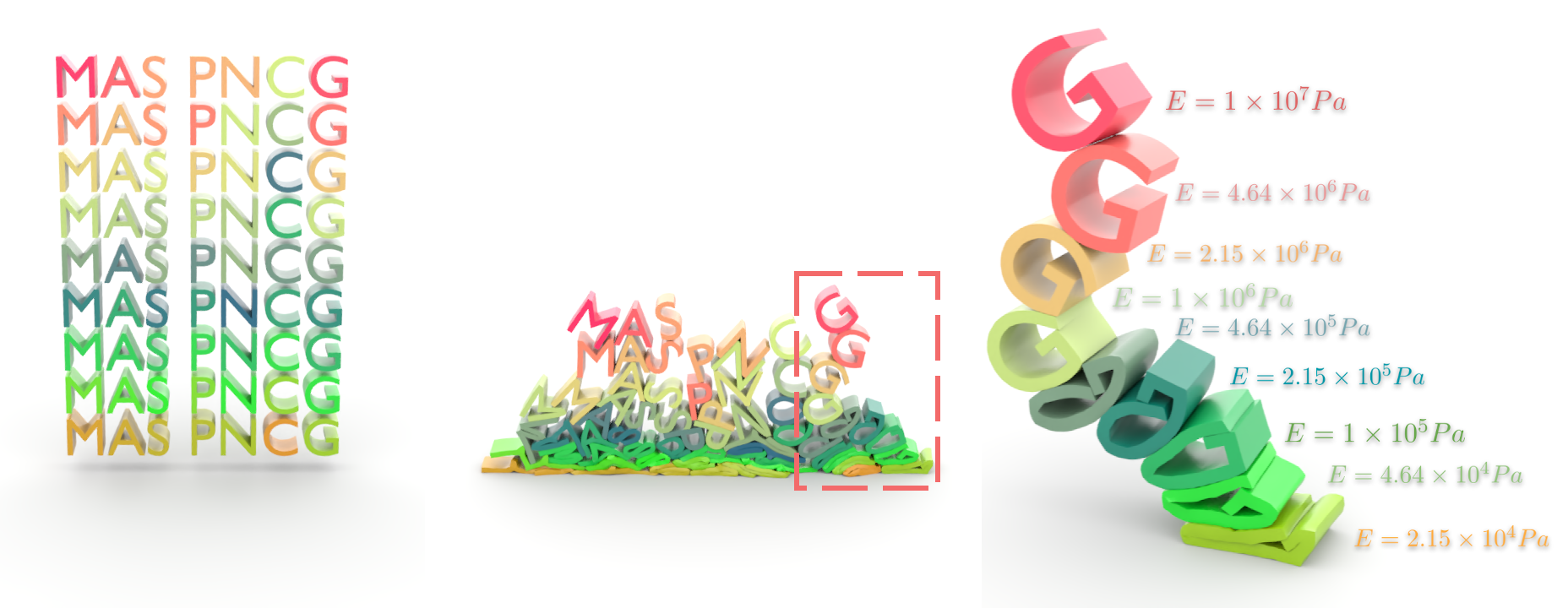}
\vspace{-10pt} 

\caption{\textbf{Dropping Letters}. A stack of 3D letters whose Young’s moduli grow geometrically from $2.15\times10^{4}\,\mathrm{Pa}$ (bottom) to $1\times10^{7}\,\mathrm{Pa}$ (top).  
Left: the initial frame.
Center: a frame during the simulation.
Right: a zoom-in view of the dashed box in the center figure.
MAS-PNCG remains robust across this $500\times$ stiffness span, whereas Jacobi frequently fails to converge.
\imgref{pics:convergence_stiff} shows the convergence comparison of different stiffness.}
\vspace{-10pt} 

\label{pics:sig_asia}
\end{figure}

The ``Dropping Letters’’ scene (\imgref{pics:sig_asia}) is designed to challenge the solver with extreme variations in material stiffness. We utilize the SNH model for a stack of 3D letters whose Young’s modulus increases geometrically from $2.15\times10^{4}\,\mathrm{Pa}$ (bottom) to $1\times10^{7}\,\mathrm{Pa}$ (top), resulting in an approximately 500-fold stiffness span. Despite this extreme three–order-of-magnitude stiffness disparity, MAS-PNCG converges robustly, whereas Jacobi-PNCG frequently stalls and fails to converge. This performance difference is further illustrated by a convergence comparison across varying stiffness levels in \imgref{pics:convergence_stiff}. This comparison confirms that MAS-PNCG maintains robust convergence even with highly stiff materials, whereas the convergence rate of Jacobi-PNCG degrades severely.

\begin{figure}[htbp]
    \centering
        \centering 
        \includegraphics[width=0.9\linewidth]{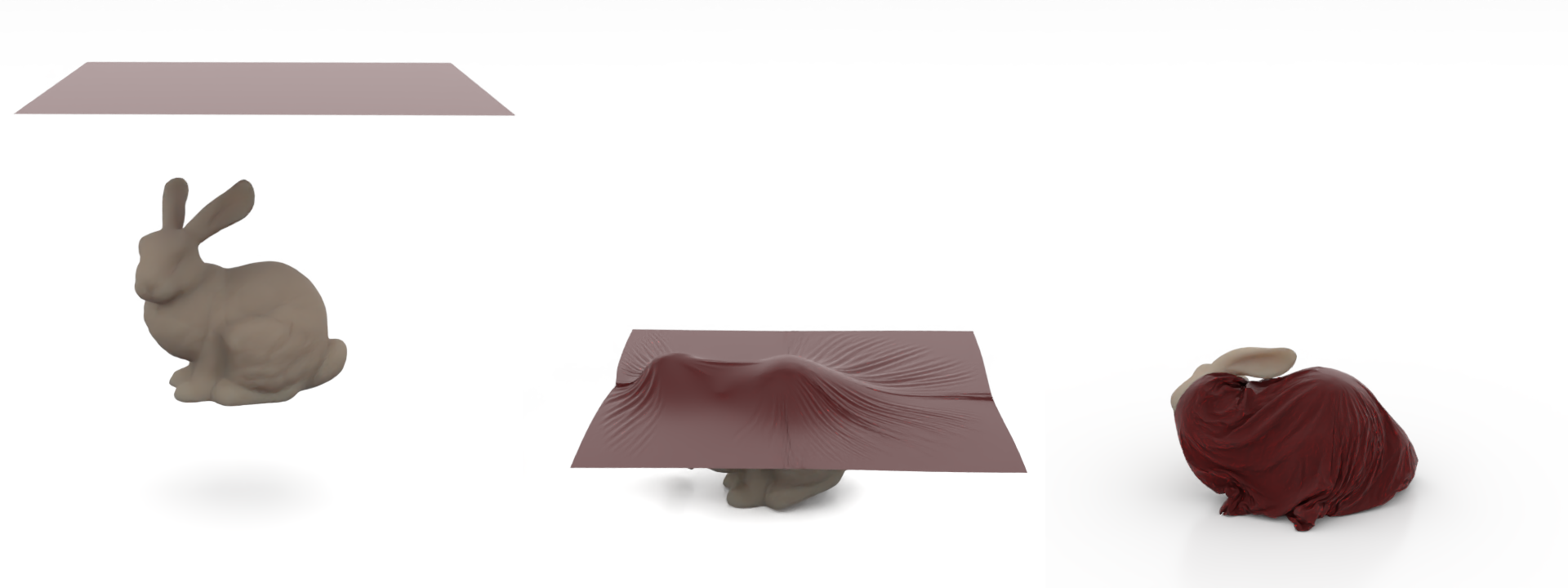}
        \vspace{-10pt} 

        \captionof{figure}{\textbf{Bunny Cloth.} A cloth sheet falls onto a dynamic bunny with friction.  MAS-PCG remains stable and faithfully captures friction-induced wrinkles and sliding.
        } 
        \label{pics:bunny_cloth}
\end{figure}

We also evaluate robustness under frictional contact. In the ``Bunny Cloth'' scene (\imgref{pics:bunny_cloth}), we simulate a cloth sheet falling onto a bunny with frictional contact. The friction model we used is same with ~\cite{li2020incremental}. MAS-PCG remains stable and accurately captures friction-induced wrinkles and sliding behavior. 
\vspace{-5pt}
\subsubsection*{Preconditioner Update Strategy.}

\begin{figure}[htbp!]
\centering
\begin{subfigure}[b]{0.4\textwidth}
    \centering
    \includegraphics[width=\columnwidth, trim=0 0 0 0, clip]{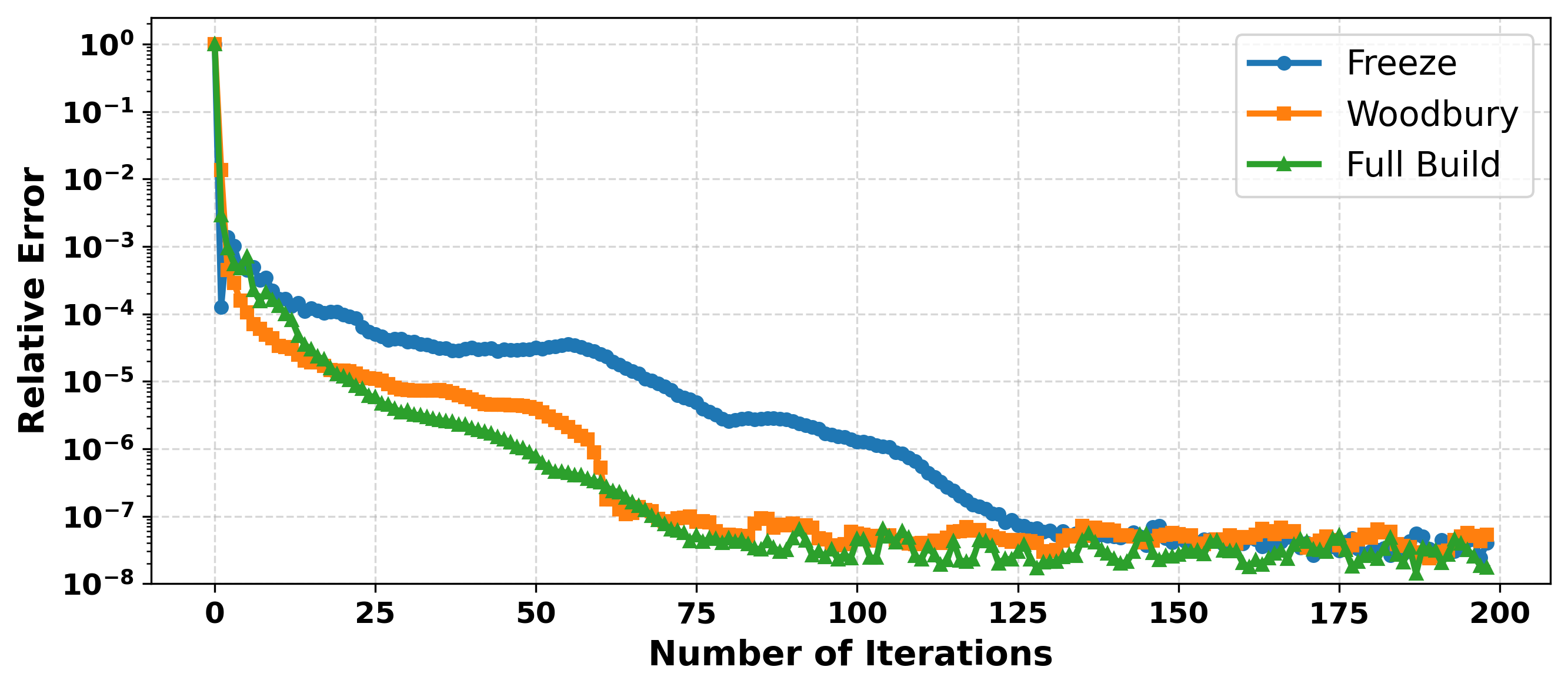}
\end{subfigure}
\vspace{-10pt} 
\caption{Comparison of preconditioner updating strategies (the "Octopus Stack" scene). Our Sparse-Input Woodbury updating achieves convergence rates comparable to \emph{Full Rebuild}, while converging substantially faster than the \emph{Freeze}.}
\label{pics:restart}
\end{figure}

We compare three preconditioner update strategies in the ``Octopus Stack'' scene (\imgref{pics:octopus_stack}):
(i) \emph{Full Rebuild}: recomputes all hierarchy levels at every PNCG iteration;
(ii) \emph{Freeze}: keeps the preconditioner fixed between NCG restarts; and
(iii) our proposed \emph{Sparse-Input Woodbury update}.
\TT{A critical question is whether our incremental Woodbury updates degrade the quality of the preconditioner compared to a ground-truth Full Rebuild. \imgref{pics:restart} demonstrates that Sparse-Input Woodbury updates preserve the convergence rate of a Full Rebuild almost exactly, yet they reduce the computation time by two orders of magnitude.
This confirms that our method successfully captures the spectral changes of the contact system without the computational penalty of explicit assembly. The \emph{Freeze} strategy, which ignores these updates, fails to converge efficiently, proving that tracking contact evolution is mandatory for PNCG.}
Since NCG restarts occur every $3$–$10$ PNCG iterations across all test scenes, the computational savings from lightweight Woodbury updates accumulate substantially.

\begin{figure}[htbp]
\centering
\begin{subfigure}[t]{0.4\textwidth}
 \centering 
\includegraphics[width=\textwidth]{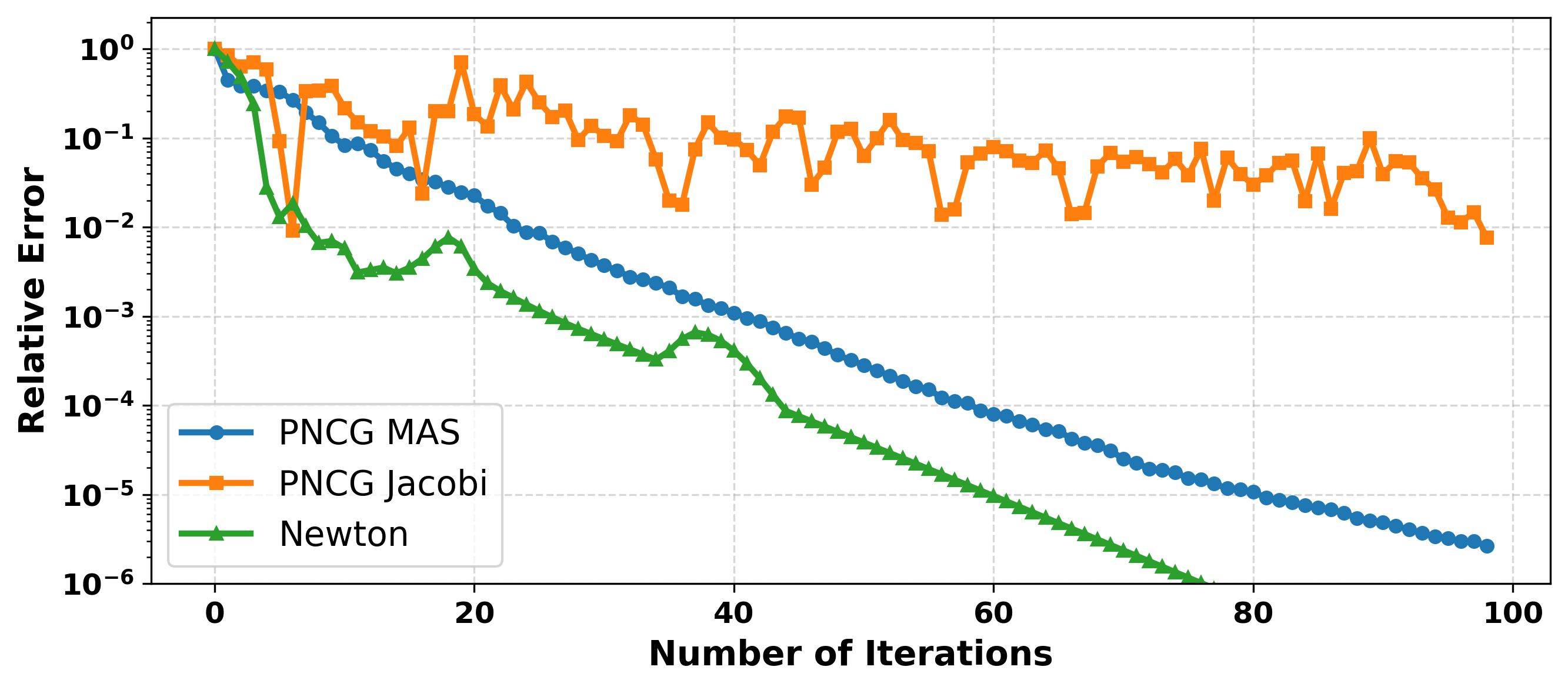}
        \caption{}
\label{pics:convergence_speed}
\end{subfigure}
\begin{subfigure}[t]{0.4\textwidth}
 \centering 
\includegraphics[width=\linewidth]{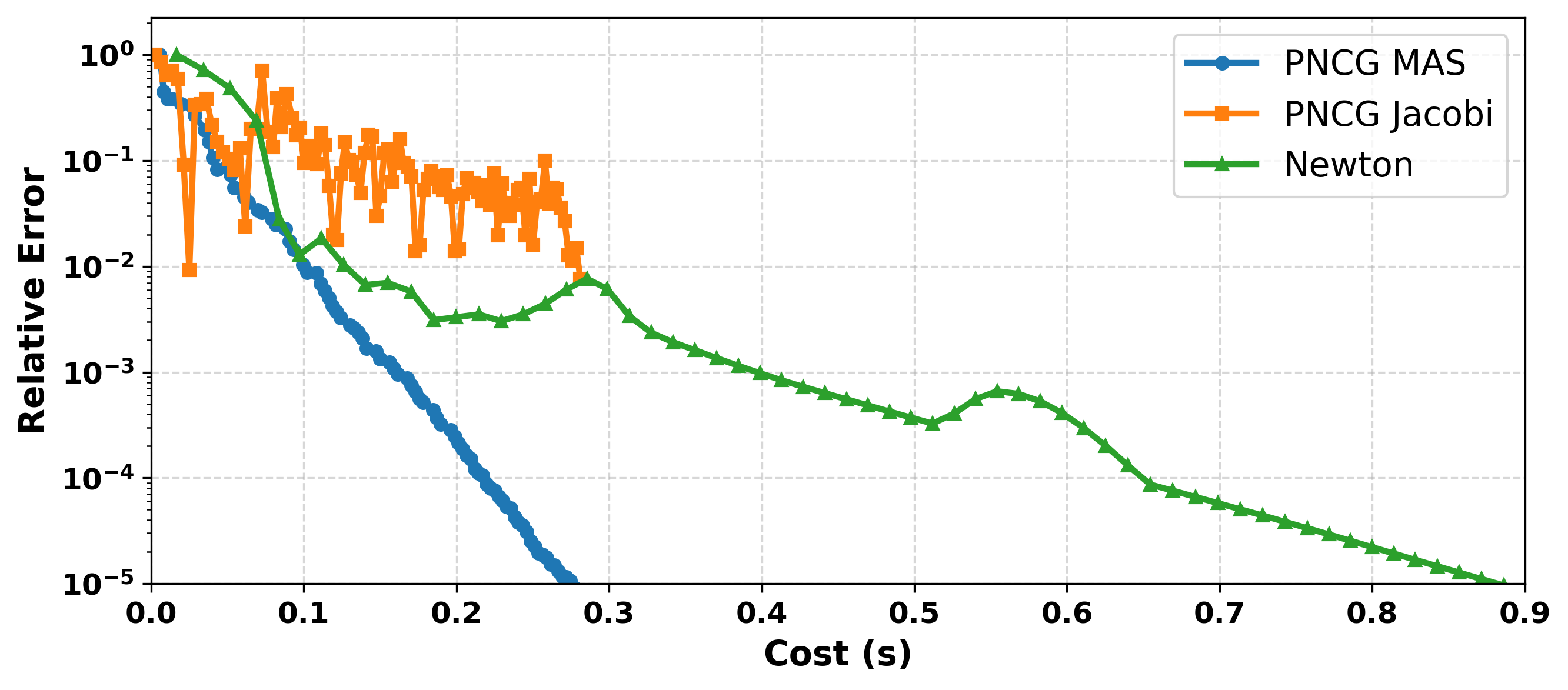}
\caption{}
\label{pics:convergence_time}
\end{subfigure}
\vspace{-10pt} 
\caption{Convergence and performance comparison between MAS-PNCG, Jacobi-PNCG, and GPU Newton-PCG with MAS preconditioner (GIPC). a) shows the relative error curve with the number of iterations. MAS preconditioning enables significantly higher accuracy convergence than Jacobi-PNCG. b) shows the relative error curve over time. MAS-PNCG reaches target accuracy faster due to reduced per-iteration overhead.} 
\vspace{-10pt}

\label{pics:convergence}
\end{figure}

We further compare MAS-PNCG with Jacobi-PNCG and the SOTA GPU Newton–PCG solver with MAS preconditioner (GIPC)~\cite{huang2024gipc}.  As shown in \imgref{pics:convergence}, MAS-PNCG reaches significantly lower error levels than Jacobi-PNCG. Although Newton-PCG converges faster per iteration, MAS-PNCG achieves the same error in less total time due to its substantially lower per-iteration cost.

\subsubsection*{Conservative CCD}
\begin{figure}[htbp]
\centering
\begin{subfigure}[b]{0.4\textwidth}
    \centering
    \includegraphics[width=\columnwidth, trim=0 0 0 0, clip]{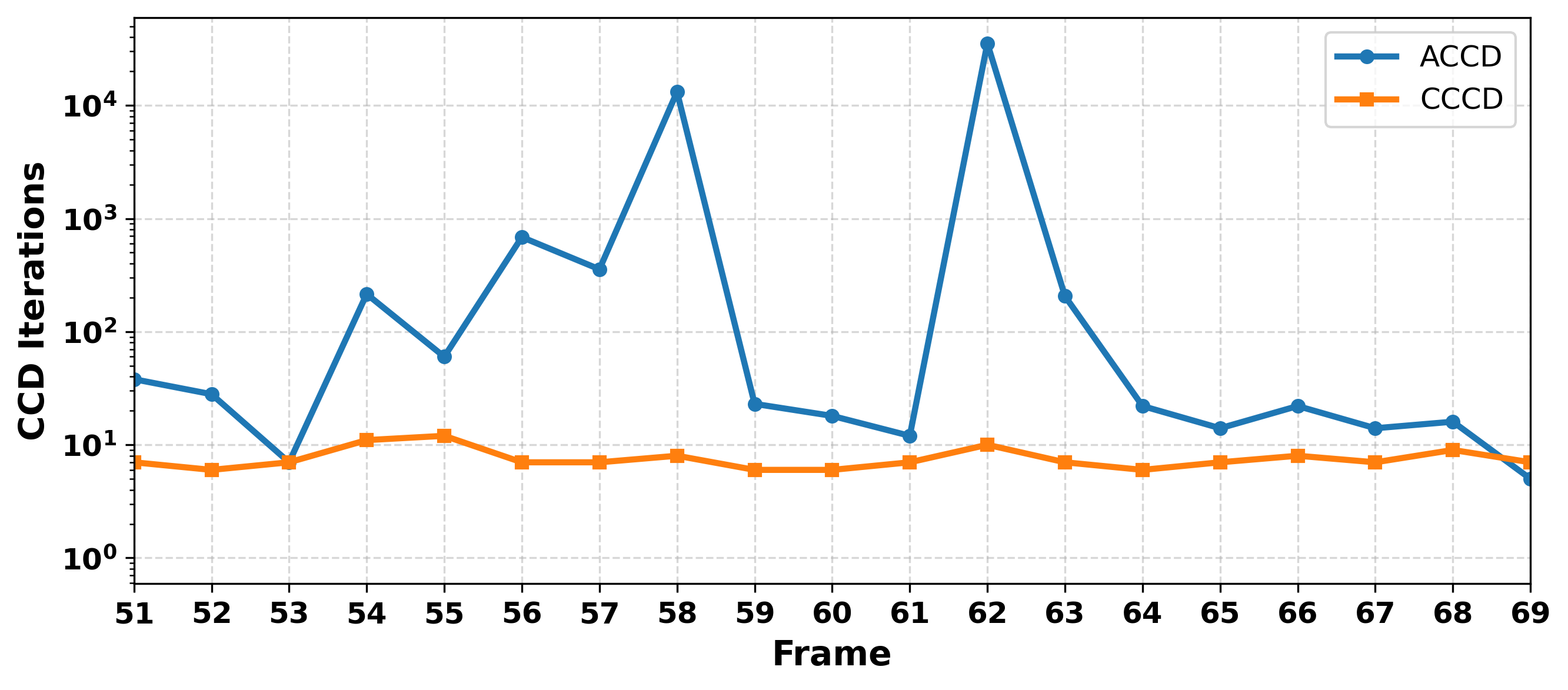}
\end{subfigure}
\vspace{-10pt} 
\caption{Comparison of CCD iteration counts per NCG step between CCCD and ACCD methods across frames following \imgref{pics:octopus_stack} (left). CCCD dramatically reduces iterations (tens vs. 30,000+) while maintaining penetration-free guarantees by prioritizing safe steps over exact TOI computation, resulting in approximately 10\% end-to-end performance improvement.}
\vspace{-5pt} 
\label{pics:ccd}
\end{figure}

\begin{figure}[htbp]
    \centering
        \includegraphics[width=0.45\textwidth, trim=0 70 0 0, clip]{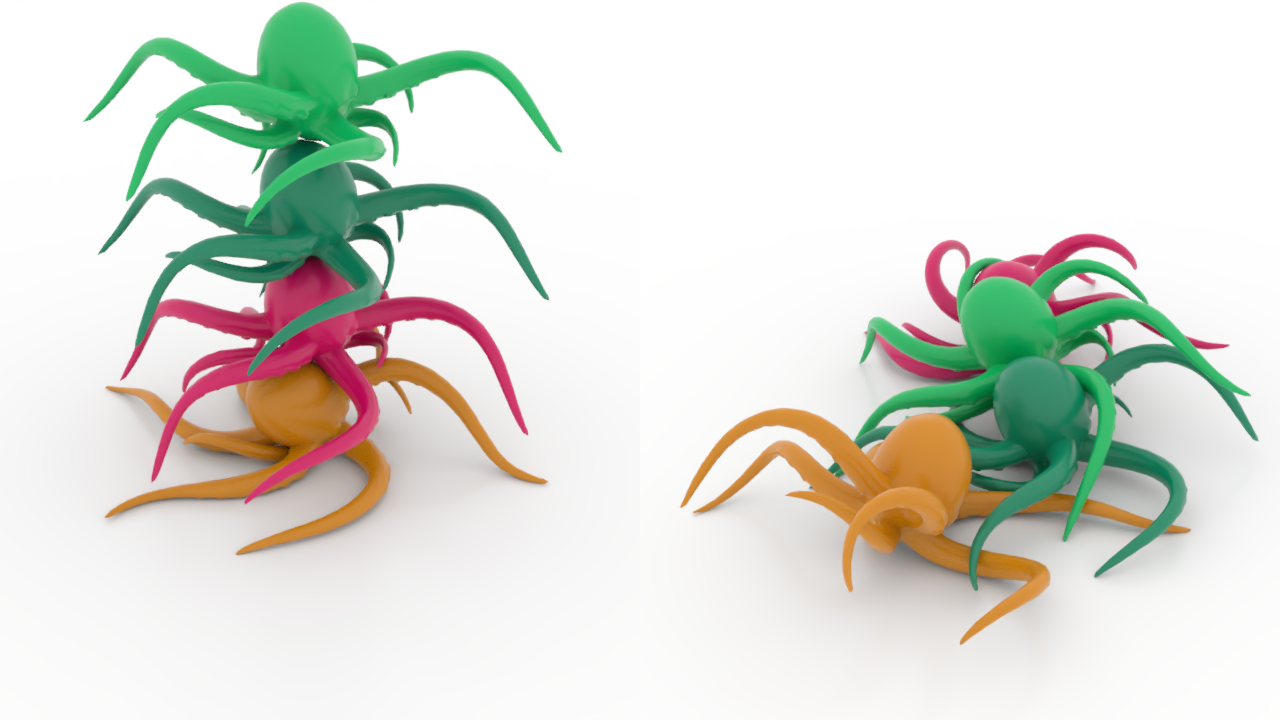}
        \vspace{-10pt} 
        \caption{\textbf{Octopus Stack.} Four octopuses fall to the ground. The frame on the left shows that the octopuses first come into mutual contact. We record the average \emph{maximum} number of CCD iterations required per NCG step over several successive frames. The results are plotted in \imgref{pics:ccd}.
        }
        \label{pics:octopus_stack}

\end{figure}

We compare our proposed Conservative Continuous Collision Detection (CCCD) with the existing Additive CCD (ACCD) method using the ``Octopus Stack'' scene.
At the frame (left) shown in \imgref{pics:octopus_stack} (approximately frame~60, when the octopuses first come into mutual contact) we record the average \emph{maximum} number of CCD iterations required per NCG step over several successive frames.  
\imgref{pics:ccd} shows that CCCD dramatically reduces iteration count from over 30,000 (ACCD) to just a few dozen, while still ensuring penetration-free configurations.
This reduced CCD computational burden improves stability within each NCG iteration and achieves an end-to-end speedup of $ \sim 10 \%$ for this test case.

\begin{figure}[htbp]
    \centering
        \centering 
        \includegraphics[width=0.4\textwidth]{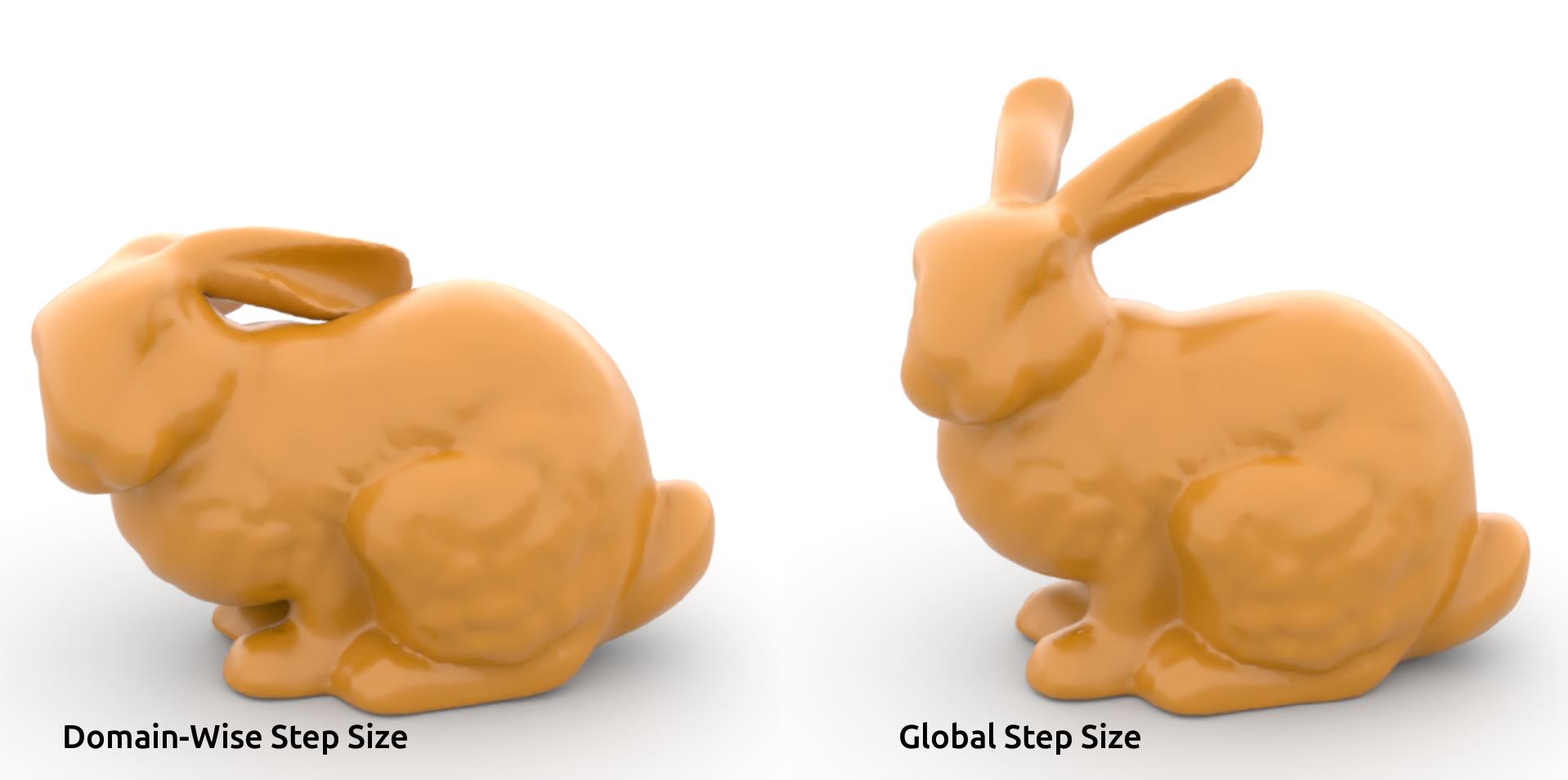}
        \vspace{-10pt} 
        \captionof{figure}{\textbf{Single Bunny.}  A bunny falls to the ground. With per-subdomain step sizes the bunny’s ears sag naturally under gravity, whereas the global strategy yields noticeably overdamped motion despite identical termination criteria.  
        }
        \label{pics:single_bunny}
  
\end{figure}

We implement CCCD on a per-subdomain basis to
avoid "numerical locking". We compare per-subdomain and global step-size selection on the ``Single Bunny’’ scene.
\imgref{pics:single_bunny} displays the same simulation frame for both strategies.  
Per-subdomain step sizes allow the bunny's ears to sag naturally under gravity, while the global strategy produces visibly overdamped motion even with identical convergence criteria. The global approach requires more than five times the time costing to achieve comparable visual quality.

\vspace{-5pt}



\subsubsection*{Performance}

Table~\ref{tab:performance_comparison_gipc} presents a detailed performance comparison between MAS-PNCG and two SOTA Newton-based solvers: GIPC and StiffGIPC.
\TT{Crutially, both baseslines already employ MAS preconditioning within their inner loops. This highlights that our performance advantage does not come from MAS itself, but from how MAS is applied. In both baselines, the high cost the Newton outer loop (e.g. Hessian rebuilding) bounds performance. By shifting to a NCG framework equipped with our our Sparse-Input Woodbury maintainance scheme, we eliminate the assembly bottleneck while retaining the spectral benefits of MAS. Also, our Hessian-aware 2D subspace minimization ensures optimal descent directions, preventing the convergence degradation typically associated with standard PNCG.}

We break down the per-frame computation into several key stages: gradient and Hessian calculation, search direction computation, CCD and line search.

An observation is the dramatic reduction in search direction computation time. While GIPC and StiffGIPC spend significant time solving linear systems (0.13--7.9s depending on scene complexity), our method computes preconditioned search directions in just 0.01--0.39s by avoiding explicit Hessian assembly and leveraging efficient MAS preconditioning with Woodbury updates. This stage contributes the majority of our speedup.

Another observation is that while our method substantially reduces the bottleneck time in computing search directions, the quasi-Newton nature of PNCG requires more iterations to converge compared to full Newton methods. Consequently, the cumulative time spent on CCD and line search increases proportionally with the iteration count. Nevertheless, the dramatic savings in per-iteration direction computation far outweigh this overhead, resulting in net performance gains across all tested scenarios.

\begin{figure}[htbp]
        \centering 
        \includegraphics[width=1.1\linewidth, trim=0 70 0 240, clip]{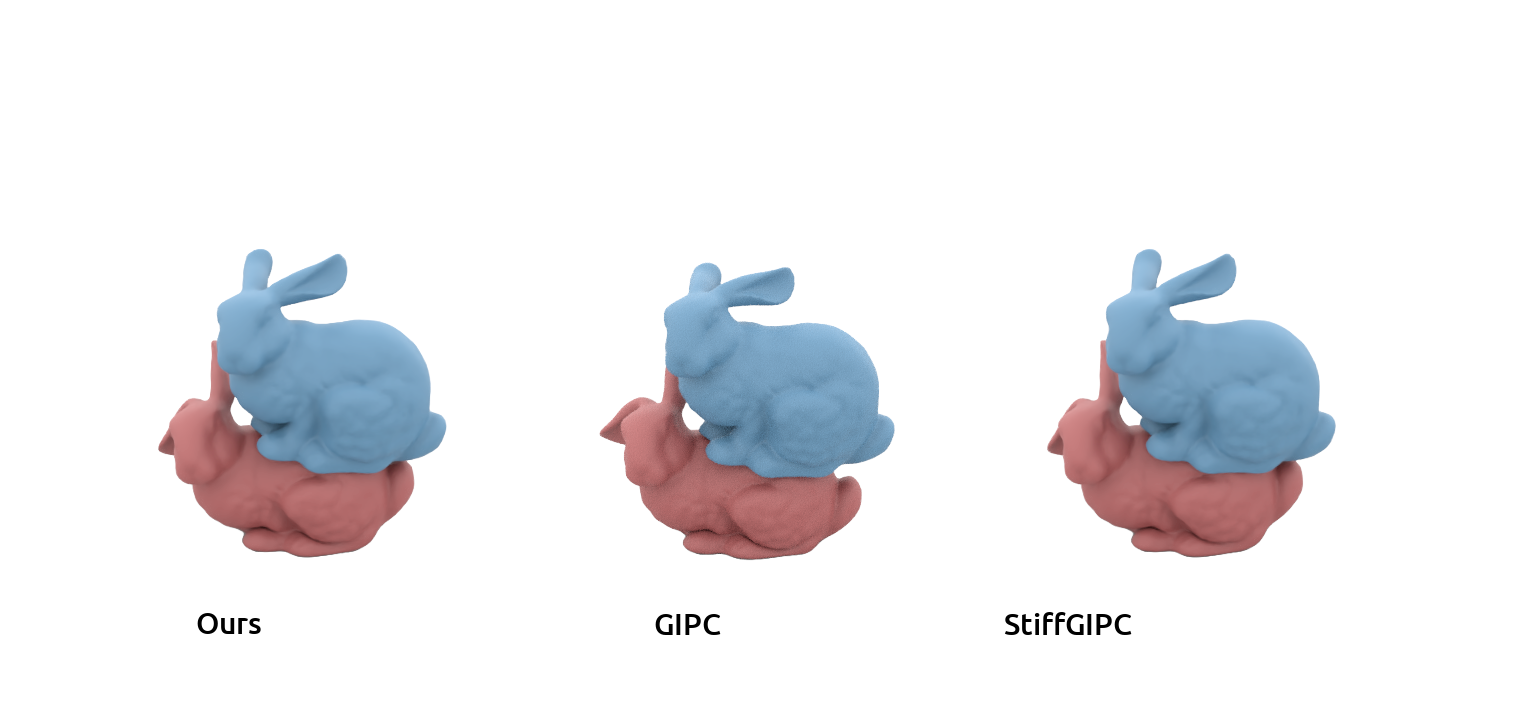}
        \vspace{-20pt} 
        \caption{\textbf{Two bunnies.} One bunny falls on top of another bunny. Our method runs in $0.21$ s per frame compared with GIPC’s $1.18$ s and StiffGIPC's $0.43$ s. Despite this substantial runtime reduction, the visual results obtained by our faster MAS-PNCG solver are 	virtually comparable to those produced by the full Newton-based GIPC method.} 
        \label{pics:two_bunnies}
\end{figure}

Across five benchmark scenes with varying scales (8K--78K vertices), MAS-PNCG consistently outperforms both baselines. We achieve 2.12--5.66$\times$ speedup over GIPC and 1.03--2.07$\times$ speedup over StiffGIPC. The largest gains occur in the scenario ``Two Bunnies'' (5.66$\times$ over GIPC, 2.07$\times$ over StiffGIPC. See \imgref{pics:two_bunnies}) where our incremental preconditioner updates efficiently track evolving contact configurations. Even in the challenging ``Dragon High'' scene with 274K tetrahedra, we maintain a 3.25$\times$ speedup over GIPC while achieving comparable accuracy.

These results confirm that our method's efficiency advantage is not simply due to using a different solver framework, but rather stems from our key algorithmic innovations: (1) Sparse-Input Woodbury updates that exploit the low-rank structure of contact Hessian changes for efficient preconditioner maintenance, and (2) 2D subspace minimization that leverages curvature information for optimal search directions. Together, these techniques enable MAS-PNCG to outperform Newton-PCG methods that already use the same MAS preconditioning.


\subsection{Stress Tests}
To further validate robustness and scalability under extreme conditions we ran several highly demanding benchmarks.
\vspace{-2pt}
\subsubsection*{Stretching Armadillo} 
\begin{figure}[htbp]
\centering
\begin{subfigure}[t]{0.4\textwidth}
\includegraphics[width=\textwidth, trim=0 0 0 0, clip]{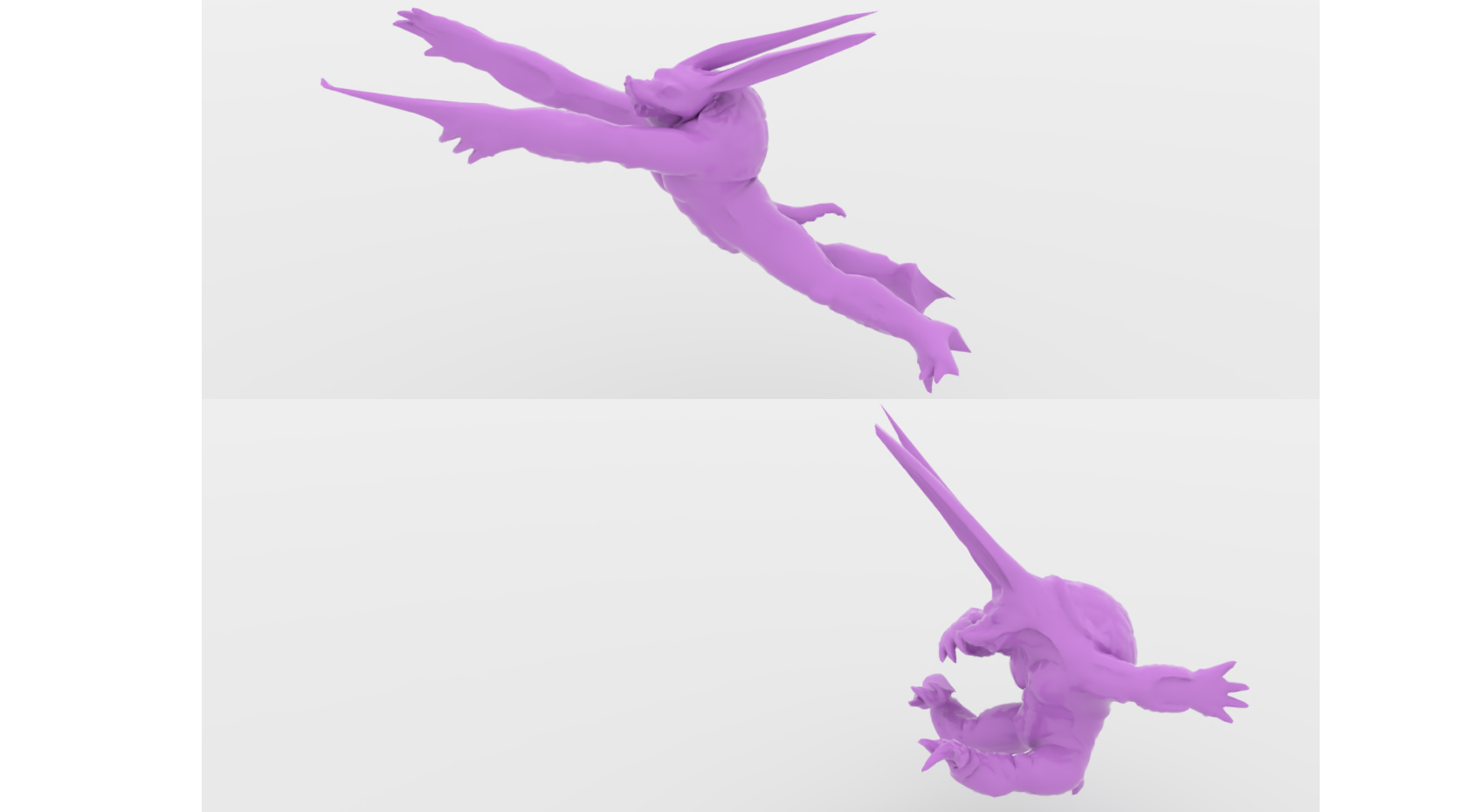}
\end{subfigure}

\caption{\textbf{Stretching Armadillo.} (1) shows armadillo is gradually stretched by a large external force. (2) shows that the armadillo rebound within a few frames when the stretching force is suddenly removed.}
\label{pics:armadillo}

\end{figure}

In \imgref{pics:armadillo}, we setup a large-deformation simulation scene: the armadillo is gradually stretched by a large external force, then the force is suddenly removed. The result shows that our method allows the armadillo to rebound within a few frames. This demonstrates that our method can maintain convergence comparable to that of Newton’s method in large-deformation scenarios.

\subsubsection*{Dragon High Resolution}
\begin{figure}[htbp]
\centering
\begin{subfigure}[t]{0.4\textwidth}
\includegraphics[width=\linewidth]{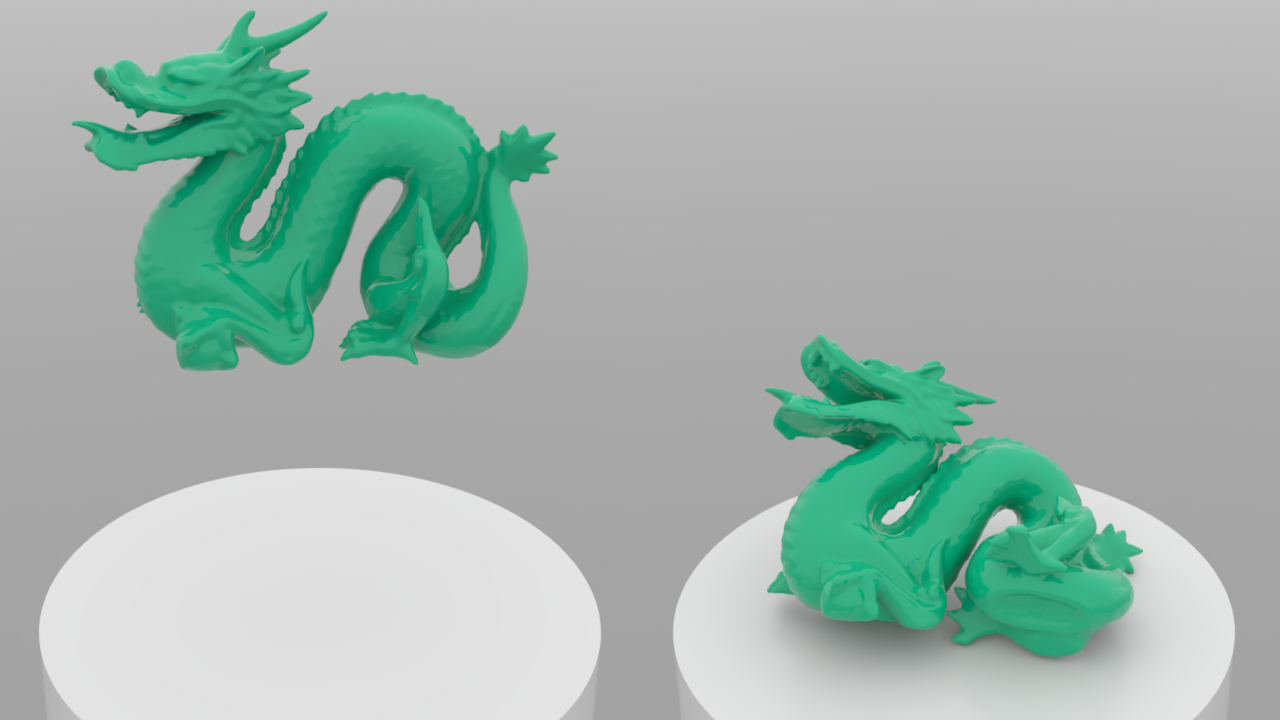}
\end{subfigure}

\caption{\textbf{Dragon High.} A free-fall scene of a high-resolution dragon model (270k tetrahedral elements).}
\label{pics:dragon}

\end{figure}

\imgref{pics:dragon} shows a free-fall scene of a high-resolution dragon model (270k tetrahedral elements). In this high-DoF setting, GIPC requires 11.05 s to simulate per frame, StiffGIPC requires 3.95 s, while our method takes only 3.39 s.

\subsubsection*{Cloth Twisting}

\begin{figure}[htbp]
\centering
\begin{subfigure}[t]{0.23\textwidth}
\includegraphics[width=\textwidth, trim=0 80 0 80, clip]{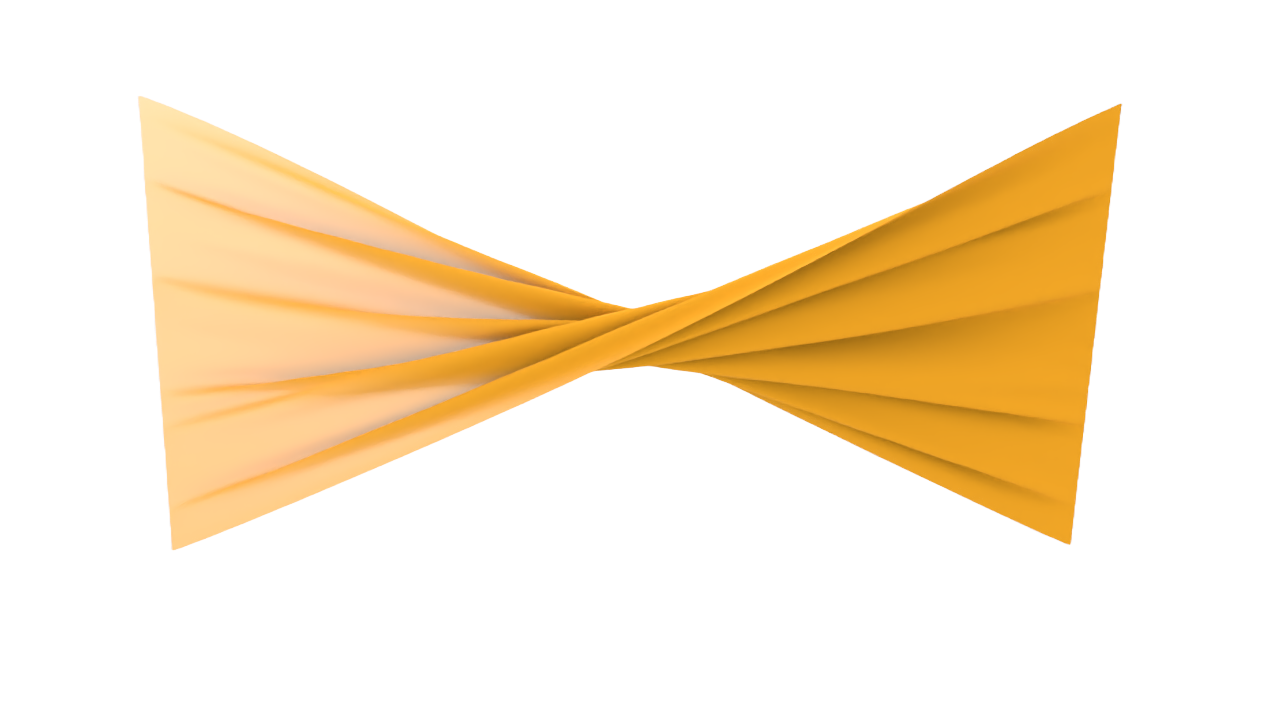}
 \vspace{-10pt} 
\end{subfigure}
\begin{subfigure}[t]{0.23\textwidth}
\includegraphics[width=\linewidth, trim=0 80 0 80, clip]{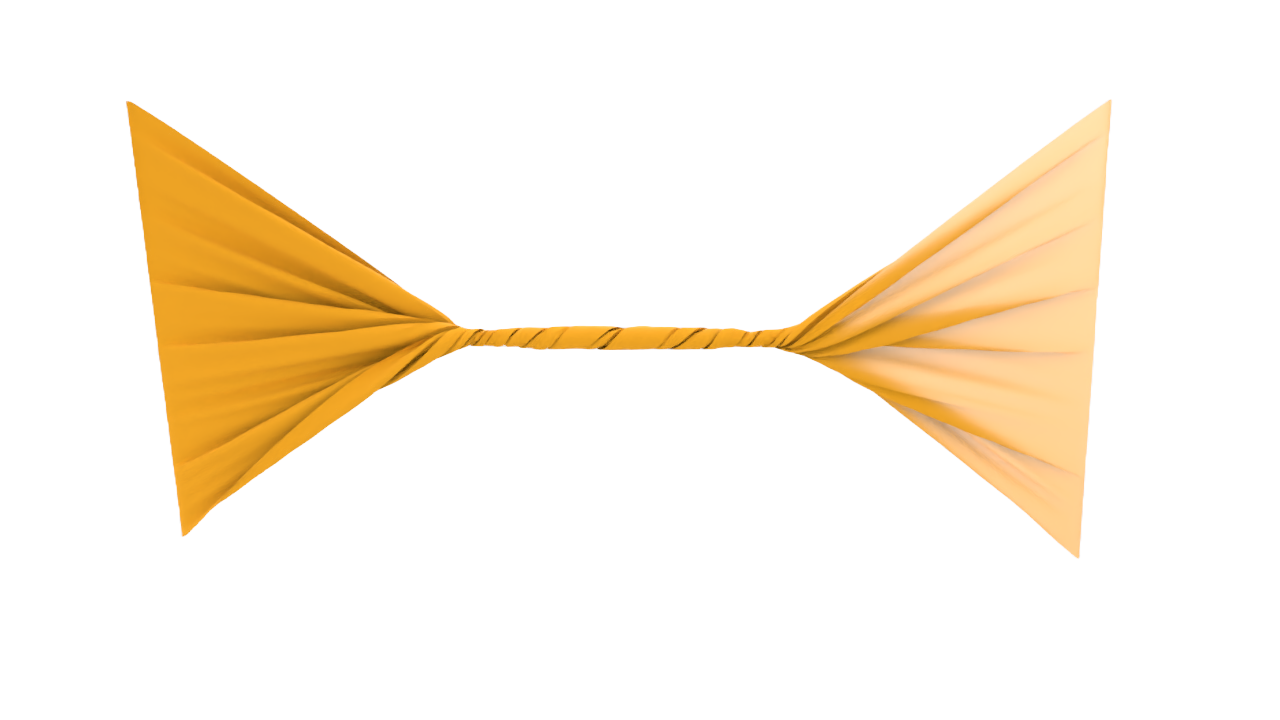}
 \vspace{-10pt} 
\end{subfigure}
 \vspace{-10pt} 
\caption{\textbf{Cloth Twisting.} A narrow cloth strip is twisted through four full turns.
Jacobi-PNCG frequently diverges or permits interpenetration in this setting.  
Conversely, MAS-PNCG converges reliably to a small error, preserves a penetration-free state.
} 
\label{pics:cloth_twisting}

\end{figure}

As shown in \imgref{pics:cloth_twisting}, we twist a narrow cloth strip through four full turns, creating dense self-collisions. Jacobi-PNCG frequently diverges or permits interpenetration in this setting. Conversely, MAS-PNCG converges reliably with low error, maintains penetration-free configurations.
  

\vspace{-2pt}
\subsubsection*{Large-Scale Simulation}

\begin{figure}[htbp]
\centering
\begin{subfigure}[t]{0.23\textwidth}
\includegraphics[width=\linewidth, trim=0 160 0 0, clip]{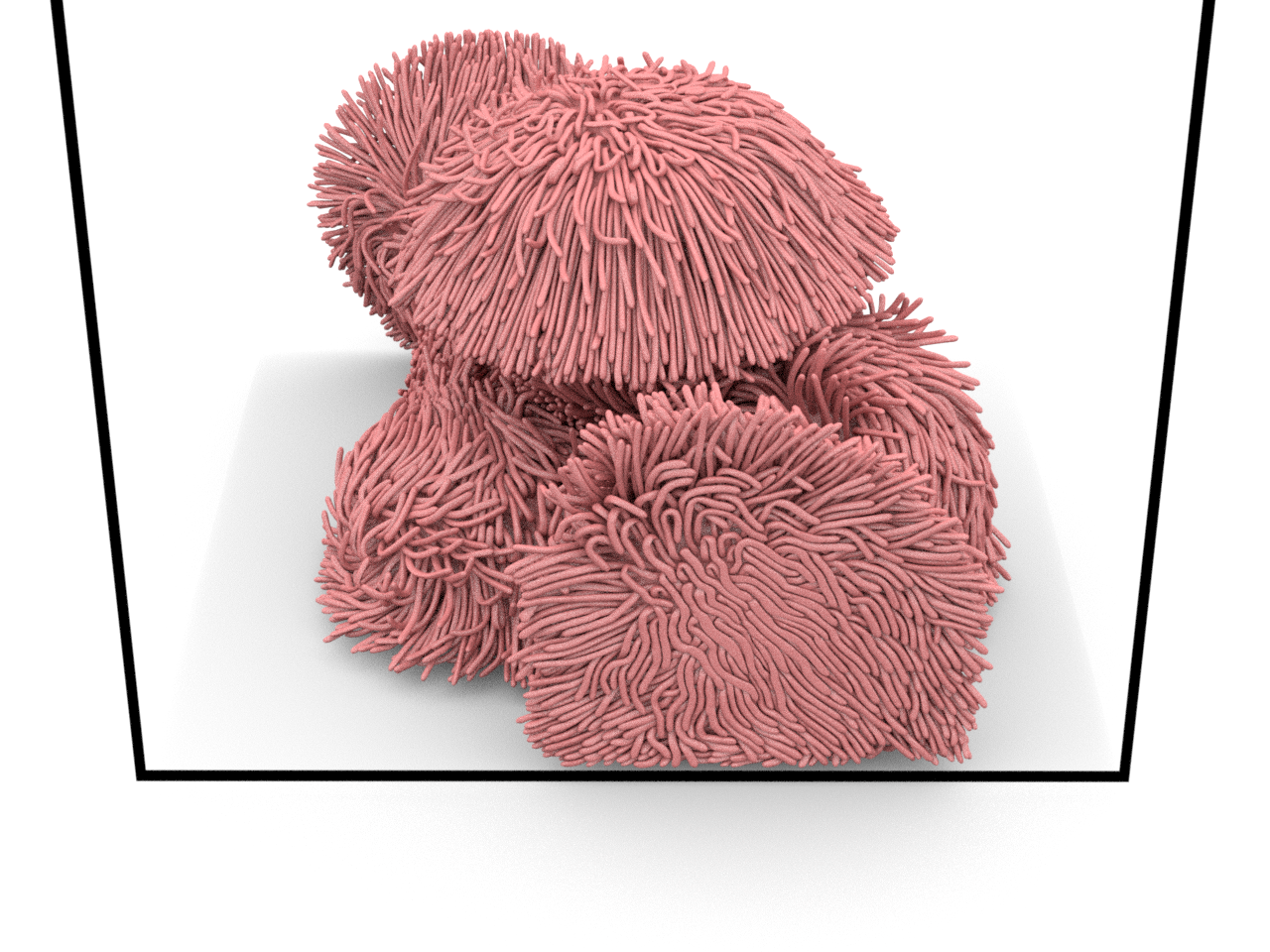}
 \vspace{-10pt} 
\end{subfigure}
\begin{subfigure}[t]{0.23\textwidth}
\includegraphics[width=\linewidth, trim=0 160 0 0, clip]{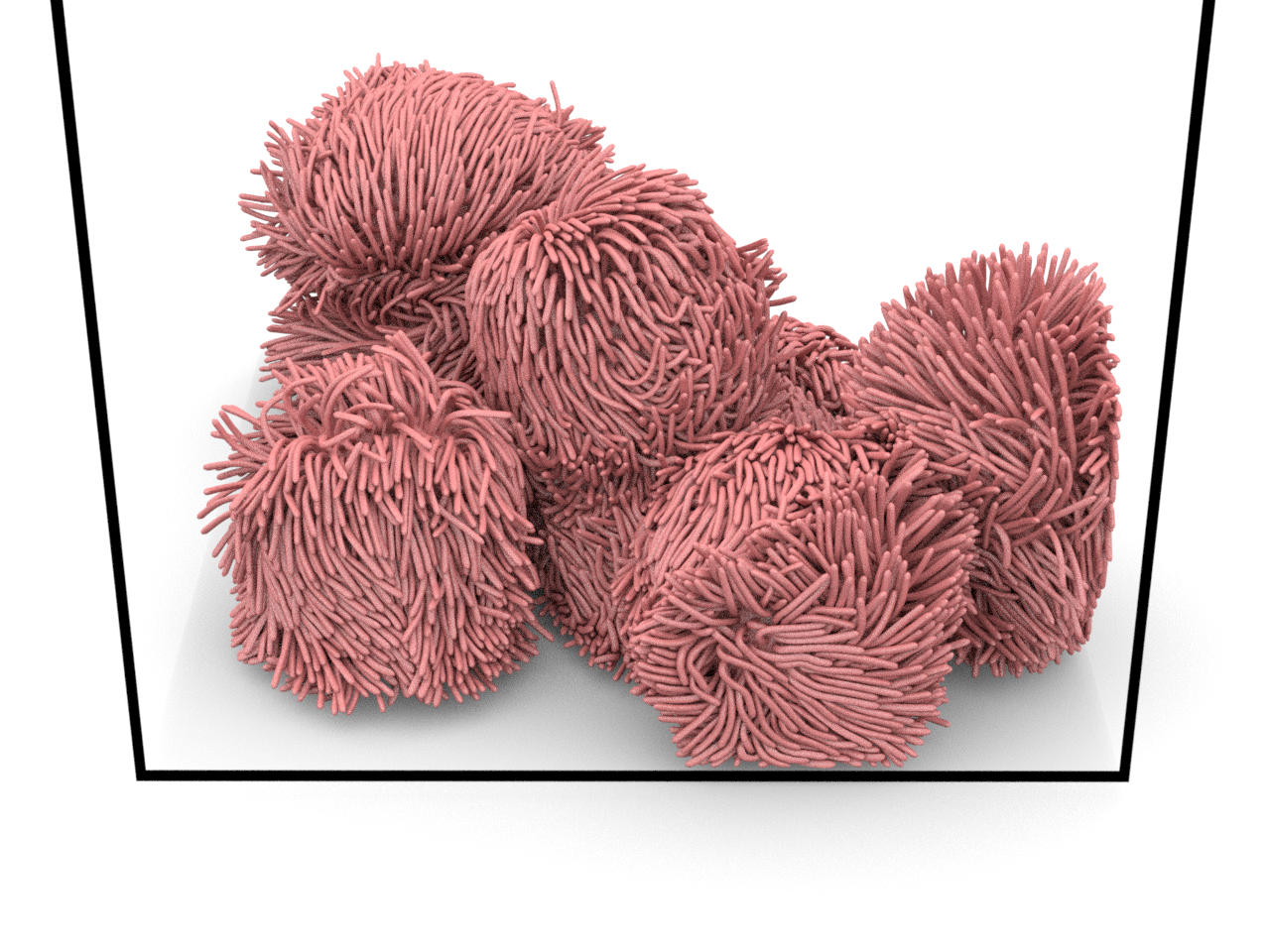}
 \vspace{-10pt} 
\end{subfigure}
 \vspace{-10pt}


\caption{\textbf{Puffer Balls.} A few frames from the same simulated scene as the \imgref{pics:teaser}.
} 
\label{pics:puffer_balls}

\end{figure}

The ``Puffer Balls'' scene (\imgref{pics:puffer_balls}) simulates 8 puffer balls with $1.14$ M vertices, $2.72$ M tetrahedra and over one million simultaneous contact pairs. Traditional Jacobi-PNCG fails due to penetrations and excessive runtime, while MAS-PNCG maintains stable, penetration-free motion and achieves convergence quality comparable to full Newton methods while remaining computationally efficient.



\section{CONCLUSION}
We have presented MAS-PNCG, an efficient and robust multilevel preconditioned nonlinear conjugate gradient method tailored for Incremental Potential Contact simulations. Our approach significantly accelerates convergence and enhances stability, especially for ill-conditioned systems arising from stiff materials or intricate contact configurations. This is achieved by integrating a Multilevel Additive Schwarz preconditioner with novel update strategies, including Sparse-Input Woodbury updates for fine-level components and Powell’s restart for global rebuilds. Complemented by an optimal 2D subspace minimization strategy for search direction and step-size determination, and a fast, conservative CCD with a tighter relative-motion lower bound for penetration-free steps, MAS-PNCG offers a compelling balance of speed and robustness.

Experimental evaluations confirm that MAS-PNCG achieves accuracy comparable to state-of-the-art MAS preconditioning Newton-PCG solvers while demonstrating substantially improved computational efficiency. This advantage is particularly pronounced in large-scale simulations, positioning our method as a scalable alternative. The ablation studies systematically validated the positive contributions of each key component: the efficient Sparse-Input Woodbury update scheme, the optimal subspace descent, the conservative CCD, and the improved lower bound formulation. We plan to release our implementation as open-source code to benefit the research community.


\bibliographystyle{ACM-Reference-Format}
\bibliography{sample-bibliography}

\clearpage

\vspace{-10pt} 

 \vspace{-10pt} 

\begin{figure}[htbp]
\centering
\begin{subfigure}[b]{0.4\textwidth}
    \centering
    \includegraphics[width=\columnwidth, trim=0 0 0 0, clip]{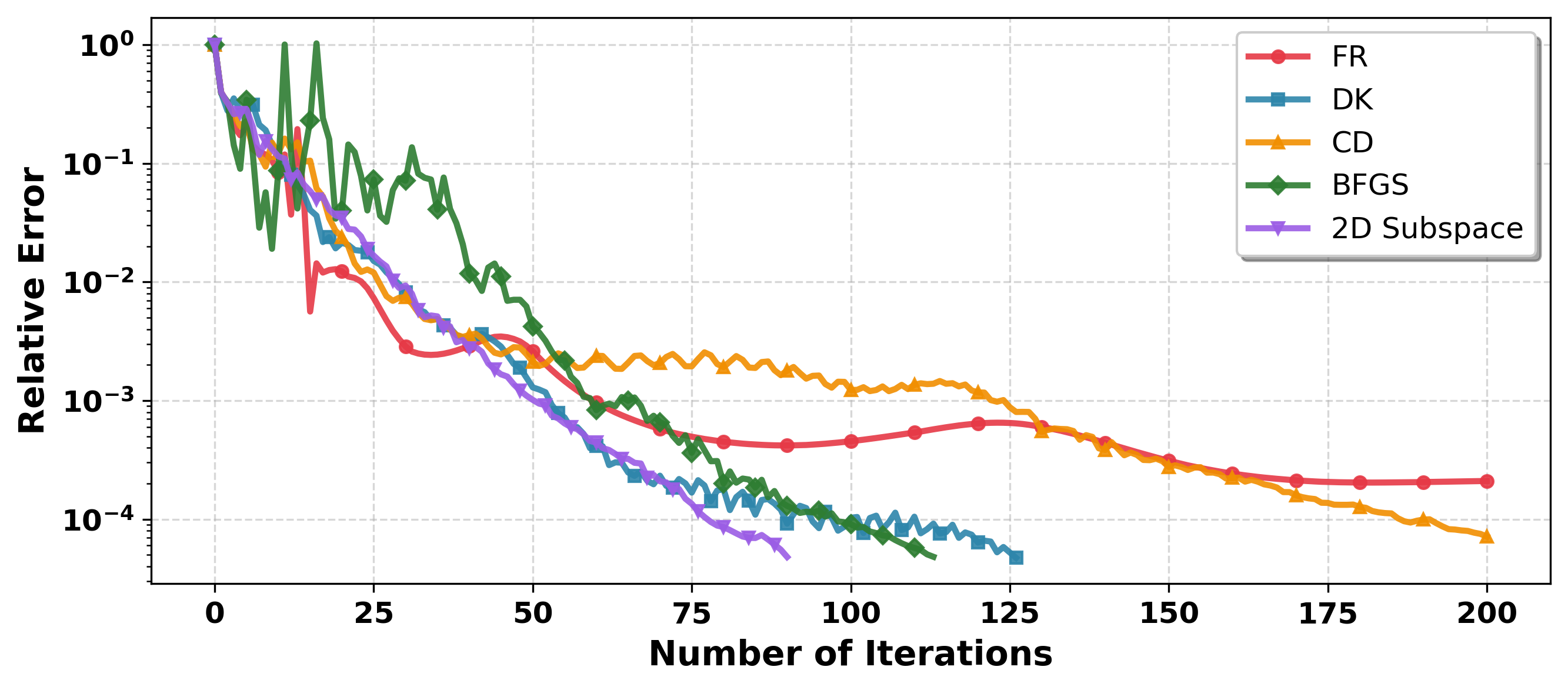}
\end{subfigure}
\vspace{-10pt} 
\caption{Convergence comparison of different nonlinear CG formulas: Fletcher-Reeves (FR), Dai-Kou (DK), Conjugate Descent (CD), BFGS, and our 2D subspace minimization. The 2D subspace method achieves consistently faster and more stable convergence.}
\vspace{-5pt} 
\label{pics:2d_subspace}
\end{figure}

\vspace{-10pt}

\vspace{-10pt}

\vspace{-10pt}

\vspace{-10pt}

\vspace{-10pt} 


 \vspace{-10pt}

 \vspace{-10pt}

\clearpage

\newpage
\clearpage


\section*{Supplementary material}

# Supplementary Document: MAS-PNCG Details and Derivations

YU ZHANG*, S-Lab, Nanyang Technological University, Singapore

XING SHEN*, Shanghai AI Laboratory, China

KEMENG HUANG†, University of Hong Kong, China

WEI CHEN, Zhejiang University, China

YIN YANG, University of Utah, United States of America

TAKU KOMURA, University of Hong Kong, China

TIANTIAN LIU, Independent Researcher, China

XINGANG PAN, S-Lab, Nanyang Technological University, Singapore

This supplementary document provides additional details for the MAS-PNCG framework. Section 1 describes the implementation of the Multilevel Additive Schwarz (MAS) preconditioner, which is crucial for handling stiff materials. Section 2 provides a detailed derivation of the $2 \times 2$ subspace minimization system used in our PNCG method (corresponding to Eq. (9) in the main paper). Section 3 provides a detailed derivation of the improved frame-invariant lower bound for the Additive Continuous Collision Detection (ACCD) step size.

## Contents

*Equal contributions.
†Corresponding author.

Authors' Contact Information: Yu Zhang, S-Lab, Nanyang Technological University, Singapore, yucrazing@gmail.com; Xing Shen, Shanghai AI Laboratory, China, shenxing@pjlab.org.cn; Kemeng Huang, University of Hong Kong, China, kmhuang@connect.hku.hk; Wei Chen, Zhejiang University, China, weichen12@zju.edu.cn; Yin Yang, University of Utah, United States of America, yangzzzy@gmail.com; Taku Komura, University of Hong Kong, China, taku@cs.hku.hk; Tiantian Liu, Independent Researcher, China, ltt1598@gmail.com; Xingang Pan, S-Lab, Nanyang Technological University, Singapore, xingang.pan@ntu.edu.sg.

## 1 Detailed Implementation of the Multilevel Additive Schwarz (MAS) Preconditioner

### 1.1 Subdomain Decomposition and Level 0 Preconditioner

The Multilevel Additive Schwarz (MAS) preconditioner is designed to handle the severe ill-conditioning inherent in Incremental Potential Contact (IPC) simulations, particularly when dealing with stiff materials and high-resolution meshes. The first step involves decomposing the computational domain $\Omega$ into a set of $D$ non-overlapping subdomains $\{\Omega_d\}_{d=1}^{D}$.

For each subdomain $\Omega_d$, we define a selection matrix $S_d \in \mathbb{R}^{3N_d \times 3N}$ that extracts the degrees of freedom (DOFs) belonging to $\Omega_d$ from the global DOF vector. The local Hessian restricted to this subdomain is given by:

$$M_{(0)}^d = S_d A S_d^T, \tag{1}$$

where $A$ is the global system matrix (Hessian of the IPC energy). The level 0 Additive Schwarz preconditioner is then defined as the sum of inverse local Hessians:

$$M_{(0)}^{-1} = \sum_{d=1}^{D} S_d^T (M_{(0)}^d)^{-1} S_d. \tag{2}$$

In our GPU implementation, each subdomain contains a fixed number of nodes, defined by a cluster size parameter (set to 16). This choice balances the degree of local coupling captured against the computational cost of batch inversion. We employ a mixed-precision strategy: the local Hessian blocks are assembled and stored in double precision to ensure numerical stability during the inversion process (using specialized high-performance batch inversion kernels), while the resulting inverted matrices are stored in single precision for high-bandwidth application during the PCG iterations.

### 1.2 Connectivity-Aware Hierarchy Construction

A critical challenge in MAS is the construction of the domain hierarchy. Traditional methods, such as those proposed in [Wu et al. 2022], often rely on Morton code-based sorting to group nodes spatially. While efficient, Morton ordering does not account for the topological connectivity of the mesh. This can lead to subdomains that contain nodes spatially close but physically disconnected (e.g., across a thin gap), resulting in poor coarse-level approximations.

Following [Huang et al. 2025], we adopt a connectivity-aware hierarchy construction. At each level $l$, we build an adjacency graph where nodes (at level 0) or clusters (at level $l > 0$) are connected if they share an element in the mesh. We then apply graph partitioning

or greedy aggregation to group these entities into larger clusters for the next level $l + 1$. Specifically, for level $l + 1$, a group of nodes from level $l$ is aggregated into a single coarse DOF. This ensures that the coarse-level basis functions represent the low-frequency modes of the elastic energy more accurately, significantly improving PCG convergence for stiff materials. In our implementation, the number of MAS levels is automatically determined based on the total number of vertices in the mesh, up to a maximum of 6 levels. Furthermore, we implement a dynamic reconnection strategy that incorporates collision-induced constraints directly into the adjacency graph. When two primitives are in close proximity, an edge is added between their corresponding nodes, ensuring that the MAS hierarchy "perceives" the physical stiffening caused by contact.

### 1.3 Restriction and Prolongation Operators

Let $N_{(l)}$ be the number of degrees of freedom at level $l$, where $N_{(0)} = 3N$. We define the coarsening (restriction) matrix $C_{(l)} \in \mathbb{R}^{N_{(l)} \times N_{(0)}}$ that maps fine-level DOFs to level $l$ DOFs. The matrix $C_{(l)}$ is a binary matrix where the $(i, j)$-th entry is 1 if fine-level DOF $j$ belongs to cluster $i$ at level $l$, and 0 otherwise.

The coarse system matrix at level $l$ is computed via the Galerkin projection:

$$A_{(l)} = C_{(l)} A C_{(l)}^{\top}. \tag{3}$$

The multilevel preconditioner $P$ is then constructed by combining the contributions from all levels:

$$P = M_{(0)}^{-1} + \sum_{l=1}^{L} C_{(l)}^{\top} M_{(l)}^{-1} C_{(l)}, \tag{4}$$

where $M_{(l)}^{-1}$ represents the Additive Schwarz preconditioner applied to the coarse system $A_{(l)}$. The mapping between levels is managed via explicit lookup tables. The application of the preconditioner follows a three-stage pipeline: (1) multi-level residual restriction, which recursively projects fine-level residuals to all coarse levels; (2) local subspace solving, which applies the local inverse matrices in parallel across all levels; and (3) correction accumulation, which aggregates the results back to the finest level. This hierarchical structure effectively addresses the system's ill-conditioning across the entire frequency spectrum: fine levels handle high-frequency local errors, while coarse levels handle low-frequency global errors.

### 1.4 GPU Implementation and Performance Optimization

To achieve high throughput on modern GPUs, we utilize warp-level primitives (such as shuffle and ballot instructions) for efficient intra-cluster communication and synchronization. The standard execution configuration uses a thread block size of 256. The use of connectivity bitmasks enables efficient management of node relationships, significantly reducing global memory traffic during the hierarchy construction and matrix assembly phases.

## 2 Derivation of the 2 × 2 Subspace Minimization System

This section provides the detailed derivation of the optimal coefficients for the search direction in our Preconditioned Nonlinear Conjugate Gradient (PNCG) method.

### 2.1 Search Direction Formulation

In our PNCG method, the search direction $p_{k+1}$ is constructed as a linear combination of the current preconditioned gradient $z_{k+1}$ and the previous search direction $p_k$:

$$p_{k+1} = -\mu z_{k+1} + \nu p_k, \tag{5}$$

where $z_{k+1} = P_{k+1} g_{k+1}$ is the preconditioned gradient, $P_{k+1}$ is the MAS preconditioner at iteration $k + 1$, and $g_{k+1}$ is the gradient at the current iterate.

### 2.2 Quadratic Model Minimization

To determine the optimal coefficients $(\mu^*, \nu^*)$, we minimize a local quadratic model of the objective function. Let $Q(p)$ be the second-order Taylor expansion of the energy function around the current iterate:

$$Q(p) = E(x_k) + g_{k+1}^{\top} p + \frac{1}{2} p^{\top} \tilde{H} p, \tag{6}$$

where $\tilde{H}$ is an approximation to the Hessian, constructed via the Sparse-Input Woodbury formulation described in the main paper.

Substituting the search direction from Eq. (5):

$$
\begin{aligned}
Q(\mu, \nu) &= g_{k+1}^{\top}(-\mu z_{k+1} + \nu p_k) + \frac{1}{2}(-\mu z_{k+1} + \nu p_k)^{\top} \tilde{H}(-\mu z_{k+1} + \nu p_k) \\
&= -\mu g_{k+1}^{\top} z_{k+1} + \nu g_{k+1}^{\top} p_k \\
&\quad + \frac{1}{2}\left(\mu^2 z_{k+1}^{\top} \tilde{H} z_{k+1} - 2\mu\nu z_{k+1}^{\top} \tilde{H} p_k + \nu^2 p_k^{\top} \tilde{H} p_k\right). \tag{7}
\end{aligned}
$$

### 2.3 Optimality Conditions

Setting the partial derivatives to zero:

$$\frac{\partial Q}{\partial \mu} = -g_{k+1}^{\top} z_{k+1} + \mu z_{k+1}^{\top} \tilde{H} z_{k+1} - \nu z_{k+1}^{\top} \tilde{H} p_k = 0, \tag{8}$$

$$\frac{\partial Q}{\partial \nu} = g_{k+1}^{\top} p_k - \mu z_{k+1}^{\top} \tilde{H} p_k + \nu p_k^{\top} \tilde{H} p_k = 0. \tag{9}$$

Rearranging these equations into matrix form:

$$\begin{bmatrix} z_{k+1}^{\top} \tilde{H} z_{k+1} & -z_{k+1}^{\top} \tilde{H} p_k \\ -p_k^{\top} \tilde{H} z_{k+1} & p_k^{\top} \tilde{H} p_k \end{bmatrix} \begin{bmatrix} \mu \\ \nu \end{bmatrix} = \begin{bmatrix} g_{k+1}^{\top} z_{k+1} \\ -g_{k+1}^{\top} p_k \end{bmatrix}. \tag{10}$$

Note that the matrix is symmetric since $z_{k+1}^{\top} \tilde{H} p_k = p_k^{\top} \tilde{H} z_{k+1}$.

### 2.4 Efficient Computation via Sparse-Input Woodbury Structure

As described in the main paper (Eq. (2)), we employ a Sparse-Input Woodbury formulation for the Hessian approximation:

$$\tilde{H} = H_{\text{base}} + UU^{\top}, \tag{11}$$

where $H_{\text{base}}$ is the frozen Hessian snapshot from the last global restart, and $U = [u_1, u_2, \ldots, u_m] \in \mathbb{R}^{n \times m}$ is the update matrix collecting rank-1 contributions $u_i = \sqrt{k_i} n_i$ from $m$ active contacts. Here $k_i = b''(d_{c_i})$ is the scalar stiffness derived from the barrier function's second derivative, and $n_i = \nabla d_{c_i}$ is the gradient direction of the distance function for contact $i$.

This structure enables efficient computation of the $2 \times 2$ matrix entries. For any vectors $a$ and $b$, the quadratic form with $\tilde{H}$ can be decomposed as:

$$a^{\top} \tilde{H} b = a^{\top} H_{\text{base}} b + a^{\top} UU^{\top} b = a^{\top} H_{\text{base}} b + (U^{\top} a)^{\top} (U^{\top} b). \tag{12}$$

Applying this to the three unique entries of the $2 \times 2$ system matrix:

$$\mathbf{z}_{k+1}^{\top} \tilde{\mathbf{H}} \mathbf{z}_{k+1} = \mathbf{z}_{k+1}^{\top} \mathbf{H}_{\text{base}} \mathbf{z}_{k+1} + \|\mathbf{U}^{\top} \mathbf{z}_{k+1}\|^2, \tag{13}$$

$$\mathbf{z}_{k+1}^{\top} \tilde{\mathbf{H}} \mathbf{p}_k = \mathbf{z}_{k+1}^{\top} \mathbf{H}_{\text{base}} \mathbf{p}_k + (\mathbf{U}^{\top} \mathbf{z}_{k+1})^{\top} (\mathbf{U}^{\top} \mathbf{p}_k), \tag{14}$$

$$\mathbf{p}_k^{\top} \tilde{\mathbf{H}} \mathbf{p}_k = \mathbf{p}_k^{\top} \mathbf{H}_{\text{base}} \mathbf{p}_k + \|\mathbf{U}^{\top} \mathbf{p}_k\|^2. \tag{15}$$

*2.4.1 Computational Procedure.* In practice, the computation proceeds as follows:

(1) **Compute low-dimensional projections**: Calculate $\boldsymbol{\xi}_z = \mathbf{U}^{\top} \mathbf{z}_{k+1} \in \mathbb{R}^m$ and $\boldsymbol{\xi}_p = \mathbf{U}^{\top} \mathbf{p}_k \in \mathbb{R}^m$. Since $\mathbf{U}$ is sparse (each column $\mathbf{u}_i$ has only $O(1)$ nonzeros corresponding to the nodes involved in contact $i$), this costs $O(m)$.

(2) **Compute base Hessian contributions**: Only two matrix-vector products are required: $\mathbf{v}_z = \mathbf{H}_{\text{base}} \mathbf{z}_{k+1}$ and $\mathbf{v}_p = \mathbf{H}_{\text{base}} \mathbf{p}_k$. The three quadratic form terms are then computed via inner products: $\mathbf{z}_{k+1}^{\top} \mathbf{H}_{\text{base}} \mathbf{z}_{k+1} = \mathbf{z}_{k+1}^{\top} \mathbf{v}_z$, $\mathbf{z}_{k+1}^{\top} \mathbf{H}_{\text{base}} \mathbf{p}_k = \mathbf{z}_{k+1}^{\top} \mathbf{v}_p$, and $\mathbf{p}_k^{\top} \mathbf{H}_{\text{base}} \mathbf{p}_k = \mathbf{p}_k^{\top} \mathbf{v}_p$. These sparse matrix-vector products can be computed efficiently using the same routines used elsewhere in the solver.

(3) **Assemble the entries**: Using Eqs. (13)–(15):

$$A_{11} = \mathbf{z}_{k+1}^{\top} \mathbf{H}_{\text{base}} \mathbf{z}_{k+1} + \boldsymbol{\xi}_z^{\top} \boldsymbol{\xi}_z, \tag{16}$$

$$A_{12} = A_{21} = \mathbf{z}_{k+1}^{\top} \mathbf{H}_{\text{base}} \mathbf{p}_k + \boldsymbol{\xi}_z^{\top} \boldsymbol{\xi}_p, \tag{17}$$

$$A_{22} = \mathbf{p}_k^{\top} \mathbf{H}_{\text{base}} \mathbf{p}_k + \boldsymbol{\xi}_p^{\top} \boldsymbol{\xi}_p. \tag{18}$$

(4) **Solve the $2 \times 2$ system**: The final system

$$\begin{bmatrix} A_{11} & -A_{12} \\ -A_{21} & A_{22} \end{bmatrix} \begin{bmatrix} \mu \\ \nu \end{bmatrix} = \begin{bmatrix} \mathbf{g}_{k+1}^{\top} \mathbf{z}_{k+1} \\ -\mathbf{g}_{k+1}^{\top} \mathbf{p}_k \end{bmatrix} \tag{19}$$

is solved via direct inversion in $O(1)$ operations.

*2.4.2 Complexity Analysis.* The dominant cost is the sparse matrix-vector products with $\mathbf{H}_{\text{base}}$, which scale as $O(\text{nnz}(\mathbf{H}_{\text{base}}))$ where nnz denotes the number of nonzeros. However, in our PNCG framework, these products are often already computed as part of the gradient evaluation or preconditioner application, allowing them to be reused. The additional overhead from the low-rank update terms is only $O(m)$, where $m$ is the number of active contacts—typically much smaller than the system size $n$.

*2.4.3 Handling Singular or Near-Singular Cases.* The $2 \times 2$ system matrix

$$\mathbf{A}_{\text{sub}} = \begin{bmatrix} A_{11} & -A_{12} \\ -A_{12} & A_{22} \end{bmatrix} \tag{20}$$

may become singular or near-singular under certain conditions. Let $\Delta = A_{11}A_{22} - A_{12}^2$ denote its determinant. The matrix is singular when $\Delta = 0$, which occurs when:

(1) **Linear dependence**: The vectors $\mathbf{z}_{k+1}$ and $\mathbf{p}_k$ become linearly dependent, i.e., $\mathbf{p}_k \approx c \mathbf{z}_{k+1}$ for some scalar $c$. This can happen when the optimization trajectory becomes nearly straight or when the preconditioner quality degrades significantly.

(2) **First iteration**: At the very first iteration ($k = 0$), no previous search direction $\mathbf{p}_0$ exists, so the two-dimensional subspace degenerates to one dimension.

To ensure numerical robustness, we employ the following fallback strategy:

(1) **Singularity detection**: Before solving, we check whether $|\Delta| < \epsilon_{\text{sing}} \cdot \max(A_{11}A_{22}, \epsilon_{\text{abs}})$, where $\epsilon_{\text{sing}}$ is a relative tolerance (e.g., $10^{-12}$) and $\epsilon_{\text{abs}}$ is an absolute tolerance to handle the case when all entries are small.

(2) **Fallback to steepest descent**: If the matrix is detected as singular or near-singular, we abandon the two-dimensional subspace optimization and fall back to the preconditioned steepest descent direction:

$$\mathbf{p}_{k+1} = -\mathbf{z}_{k+1}, \quad \text{i.e., } \mu = 1, \ \nu = 0. \tag{21}$$

This is equivalent to restarting the conjugate gradient iteration from the current point.

(3) **First iteration handling**: At $k = 0$, we directly set $\mu = 1$ and $\nu = 0$ without attempting to solve the $2 \times 2$ system, since no previous direction is available.

This fallback mechanism ensures that even in degenerate cases, the algorithm always produces a valid descent direction. The preconditioned steepest descent direction $-\mathbf{z}_{k+1}$ is guaranteed to be a descent direction as long as the preconditioner $\mathbf{P}_{k+1}$ is symmetric positive definite, since:

$$\mathbf{g}_{k+1}^{\top}(-\mathbf{z}_{k+1}) = -\mathbf{g}_{k+1}^{\top} \mathbf{P}_{k+1} \mathbf{g}_{k+1} < 0. \tag{22}$$

## 2.5 Advantages of the $2 \times 2$ System

(1) **Elimination of Heuristic $\beta$ Selection**: Traditional NCG methods rely on heuristic formulas (e.g., Fletcher-Reeves, Polak-Ribière) to compute the conjugacy parameter $\beta$. Our approach directly optimizes both $\mu$ and $\nu$, eliminating the need for such heuristics.

(2) **Natural Step Size**: The coefficient $\mu$ acts as a "natural step size" that accounts for second-order curvature information. This allows us to adopt a unit initial trial step ($\alpha = 1.0$), which is only clamped when necessary to enforce non-penetration constraints via Conservative CCD.

(3) **Negligible Computational Cost**: Solving a $2 \times 2$ symmetric linear system requires only $O(1)$ operations, which is negligible compared to the cost of the preconditioner application and matrix-vector products.

## 3 Detailed Derivation of the Improved Relative-Displacement Lower Bound for ACCD

This section provides a detailed derivation of the improved frame-invariant lower bound for the Additive Continuous Collision Detection (ACCD) step size used in the MAS-PNCG framework. As discussed in the main text, the standard ACCD approach [Li et al. 2021] can be overly conservative. We show that by considering relative nodal displacements directly, we obtain a tighter bound that is independent of the global reference frame.

### 3.1 Standard ACCD Lower Bound

The standard ACCD method computes a conservative lower bound for the step size $\alpha$ for each potentially colliding pair of primitives (e.g., a point $P$ and a triangle $T$). Let $d(t)$ be the distance between

the primitives at time $t \in [0, 1]$. The goal is to find a step $\alpha$ such that $d(t) > 0$ for all $t \in [0, \alpha]$.

The standard approach uses the following bound for the denominator $l$ of the time-of-impact estimate:

$$l_{\text{original}} = \max \|\mathbf{p}_P\| + \max_{j \in \{0,1,2\}} \|\mathbf{p}_{T_j}\|, \tag{23}$$

where $\mathbf{p}_P$ and $\mathbf{p}_{T_j}$ are the search directions (displacements) of the point and the triangle vertices, respectively. This bound stems from the triangle inequality applied to the relative displacement:

$$\|\mathbf{p}_P - \mathbf{p}_T(u,v)\| \leq \|\mathbf{p}_P\| + \|\mathbf{p}_T(u,v)\| \leq \max \|\mathbf{p}_P\| + \max \|\mathbf{p}_{T_j}\|, \tag{24}$$

where $\mathbf{p}_T(u,v)$ is the displacement of a point on the triangle via barycentric interpolation.

### 3.2 Improved Frame-Invariant Bound

We aim to bound the magnitude of the relative displacement vector directly:

$$\Delta \mathbf{p}(u,v) = \mathbf{p}_P - \mathbf{p}_T(u,v). \tag{25}$$

Since the triangle vertices move linearly along their search directions, the displacement of any point on the triangle is a linear interpolation of the vertex displacements:

$$\mathbf{p}_T(u,v) = (1 - u - v)\mathbf{p}_{T_0} + u\mathbf{p}_{T_1} + v\mathbf{p}_{T_2}. \tag{26}$$

Thus, the relative displacement $\Delta \mathbf{p}(u,v)$ is also a linear function of the barycentric coordinates $(u, v)$:

$$\Delta \mathbf{p}(u,v) = \mathbf{p}_P - [(1-u-v)\mathbf{p}_{T_0} + u\mathbf{p}_{T_1} + v\mathbf{p}_{T_2}] = \sum_{j=0}^{2} w_j(\mathbf{p}_P - \mathbf{p}_{T_j}), \tag{27}$$

where $w_j$ are the barycentric weights ($w_0 = 1 - u - v, w_1 = u, w_2 = v$) satisfying $\sum w_j = 1$ and $w_j \geq 0$.

#### 3.2.1 Convexity of the Norm.
We are interested in the maximum magnitude of the relative displacement over the primitives for a Point-Triangle (PT) pair:

$$l_{\text{tight}}^{PT} = \max_{(u,v)} \|\Delta \mathbf{p}(u,v)\|. \tag{28}$$

Because the Euclidean norm $\| \cdot \|$ is a convex function and $\Delta \mathbf{p}(u,v)$ is a linear map from the coordinate space to the displacement space, the composition $f(u,v) = \|\Delta \mathbf{p}(u,v)\|$ is a convex function over the domain of the triangle.

A well-known property of convex functions is that their maximum over a compact convex set (like a triangle) is always attained at one of its extreme points (the vertices). Therefore, the specific formula for $l_{\text{tight}}$ in the PT case is:

$$l_{\text{tight}}^{PT} = \max_{j \in \{0,1,2\}} \|\mathbf{p}_P - \mathbf{p}_{T_j}\|. \tag{29}$$

#### 3.2.2 Generalization to Edge-Edge Pairs.
For an edge-edge pair with vertices $(E_{1a}, E_{1b})$ and $(E_{2a}, E_{2b})$, the relative displacement between any two points on the edges is:

$$\Delta \mathbf{p}(s,t) = [(1-s)\mathbf{p}_{E_{1a}} + s\mathbf{p}_{E_{1b}}] - [(1-t)\mathbf{p}_{E_{2a}} + t\mathbf{p}_{E_{2b}}], \tag{30}$$

which is a bilinear (and thus also linear along any axis) function of $(s, t)$. By the same convexity argument, the maximum magnitude must occur at one of the four vertex-pair combinations:

$$l_{\text{tight}}^{EE} = \max_{i \in \{a,b\}, j \in \{a,b\}} \|\mathbf{p}_{E_{1i}} - \mathbf{p}_{E_{2j}}\|. \tag{31}$$

### 3.3 Advantages

(1) **Tightness**: By the triangle inequality, $\|\mathbf{p}_i - \mathbf{p}_j\| \leq \|\mathbf{p}_i\| + \|\mathbf{p}_j\|$. Thus, $l_{\text{tight}} \leq l_{\text{original}}$ is always satisfied. In many cases, especially when primitives move in the same direction with high velocity, $l_{\text{tight}}$ is significantly smaller, leading to much larger and more efficient step sizes.

(2) **Frame Invariance**: The term $\|\mathbf{p}_i - \mathbf{p}_j\|$ only depends on the relative displacement. Adding a constant velocity vector $\mathbf{v}_c$ to all vertices (changing the reference frame) does not change the result: $\|(\mathbf{p}_i + \mathbf{v}_c) - (\mathbf{p}_j + \mathbf{v}_c)\| = \|\mathbf{p}_i - \mathbf{p}_j\|$. The original bound $l_{\text{original}}$ is sensitive to such changes.

### 3.4 Bisection-Based Conservative CCD

We introduce a bisection-based algorithm to compute tight, conservative CCD step sizes. Let $f(t) = at^3 + bt^2 + ct + d$ be the cubic polynomial characterizing coplanarity over time (signed tetrahedron volume for PT/EE pairs). Given an initial lower bound $\alpha_l = d_{\min}/l_{\text{tight}}$ from Section 3, the goal is to find the largest safe step $\alpha \in [\alpha_l, 1]$ with no root in $[0, \alpha]$.

Our key design principle is returning a *safe boundary* rather than the exact root, ensuring penetration-free guarantees under floating-point arithmetic.

#### 3.4.1 Algorithm Overview.
The algorithm partitions $[\alpha_l, 1]$ into monotonic segments using critical points of $f'(t) = 3at^2 + 2bt + c$. Based on the discriminant $\Delta = b^2 - 3ac$:

- $\Delta > 0$: Two extrema exist. We compute them via the stable formula $q = -(b + \text{sign}(b)\sqrt{\Delta})$, yielding $t_a = q/(3a)$ and $t_b = c/q$. This partitions $[\alpha_l, 1]$ into up to three monotonic segments.
- $\Delta = 0$ ($a \neq 0$): Single inflection at $t_{\inf} = -b/(3a)$; the cubic remains monotonic.
- $\Delta < 0$ or $a = 0$: Strictly monotonic; check $[\alpha_l, 1]$ directly.

For each monotonic segment, if a sign change occurs (indicating a root), we call SafeBisect (Alg. 1) to locate the safe boundary.

#### 3.4.2 Safe Boundary Bisection.
SafeBisect (Alg. 1) performs 16 fixed iterations, shrinking the interval by $2^{16} \approx 65536$. At each step, we evaluate the midpoint: if same sign as $y_s$, advance $x_s$; otherwise, shrink $x_u$. This fixed count enables GPU loop unrolling while providing $\sim 10^{-5}$ relative precision.

---

**Algorithm 1:** SafeBisect: Conservative Root Boundary

---

**Input:** Safe bound $x_s$, unsafe bound $x_u$, values $y_s, y_u$
**Output:** Conservative step size
**for** $i = 1$ **to** $16$ **do**
    $x_m \leftarrow (x_s + x_u)/2; \quad y_m \leftarrow f(x_m);$
    **if** $\text{sign}(y_s) \neq \text{sign}(y_m)$ **then** $x_u \leftarrow x_m;$
    **else** $x_s \leftarrow x_m;$
    $y_s \leftarrow y_m;;$
**end**
**return** $x_s;$

---

---

**Algorithm 2:** CUBICSAFE: Safe Step via Cubic Analysis

---

**Input:** Coefficients $[d, c, b, a]$, lower bound $\alpha_l$
**Output:** Safe step $\alpha$

**if** $\alpha_l \geq 1$ **then return** 1;
$x \leftarrow \max(0, \alpha_l)$;;
$y \leftarrow f(x)$;;
**if** $y = 0$ **then return** $x$;
$y_1 \leftarrow f(1)$;;
$\Delta \leftarrow b^2 - 3ac$;
**if** $\Delta > 0$ **then**
    $q \leftarrow -(b + \text{sign}(b)\sqrt{\Delta})$;;
    $(x_a, x_b) \leftarrow \text{sort}(q/3a, c/q)$;
    **if** $x_a \in (x, 1)$ **then**
        $y_a \leftarrow f(x_a)$;
        **if** $\text{sign}(y) \neq \text{sign}(y_a)$ **then return**
          $SAFEBISECT(x, x_a, y, y_a)$;
        $(x, y) \leftarrow (x_a, y_a)$;
    **end**
    **if** $x_b \in (x, 1)$ **then**
        $y_b \leftarrow f(x_b)$;
        **if** $\text{sign}(y) \neq \text{sign}(y_b)$ **then if** $x > \alpha_l$ **then return** $x$;
        **return** $SAFEBISECT(x, x_b, y, y_b)$;
        $(x, y) \leftarrow (x_b, y_b)$;
    **end**
**else if** $\Delta = 0$ **and** $a \neq 0$ **then**
    $x_i \leftarrow -b/(3a)$;
    **if** $x_i \in (x, 1)$ **then**
        $y_i \leftarrow f(x_i)$;
        **if** $\text{sign}(y) \neq \text{sign}(y_i)$ **then return**
          $SAFEBISECT(x, x_i, y, y_i)$;
        $(x, y) \leftarrow (x_i, y_i)$;
    **end**
**if** $\text{sign}(y) \neq \text{sign}(y_1)$ **then if** $x > \alpha_l$ **then return** $x$;
**return** $SAFEBISECT(x, 1, y, y_1)$;
**return** 1;

---

### 3.4.3 Implementation Notes.

Key details ensuring robustness: (1) the stable quadratic formula $q = -(b + \text{sign}(b)\sqrt{\Delta})$ avoids catastrophic cancellation; (2) when a sign change is detected past an already-advanced boundary ($x > \alpha_l$), we return $x$ directly as a valid safe bound; (3) the fixed 16-iteration bisection trades exact root finding for predictable GPU performance with sufficient precision ($\sim 1.5 \times 10^{-5}$).

*Numerical Robustness and Comparison.* While Newton's method or the Secant method can offer faster local convergence (quadratic or superlinear), we specifically choose the Bisection method for the root-finding in Conservative CCD due to its superior numerical robustness. First, Bisection provides a *guaranteed convergence* within a bracketed interval $[x_s, x_u]$, which is essential for handling the highly non-linear cubic polynomials in collision detection that may cause Newton's method to diverge or exit the valid $[0, 1]$ range. Second, Bisection is naturally *safe*; by always returning the boundary on the non-penetrating side ($x_s$), we provide a hard guarantee against interpenetration, whereas Newton-type methods may overshoot the root into a penetrating state. Finally, Bisection avoids reliance on derivatives, which can become near-zero or numerically unstable in degenerate configurations (e.g., near-coplanar primitives), ensuring stability where gradient-based methods typically fail.

This approach exploits cubic structure to achieve tighter bounds than linear approximations, reducing numerical locking in complex contact scenarios.



\end{document}